\documentclass[modern]{aastex63_2}
\usepackage{graphicx}

\received{November 2, 2018}
\revised{March 20, 2020}
\accepted{April 24, 2020}
\submitjournal{AJ}

\shorttitle{Pluto's Airglow Emissions}
\shortauthors{Steffl et al.}

\begin{document}

\title{Pluto's Ultraviolet Spectrum, Surface Reflectance, and Airglow
  Emissions}

\correspondingauthor{Andrew Steffl}
\email{steffl@boulder.swri.edu}

\author[0000-0002-5358-392X]{Andrew J. Steffl}
\affiliation{Southwest Research Institute \\
1050 Walnut Street, Suite 300 \\
Boulder, CO 80302 , USA}

\author{Leslie A. Young}
\affiliation{Southwest Research Institute \\
1050 Walnut Street, Suite 300 \\
Boulder, CO 80302 , USA}

\author{Darrell F. Strobel}
\affiliation{The Johns Hopkins University \\
Baltimore, MD}

\author{Joshua A. Kammer}
\affiliation{Southwest Research Institute \\
San Antonio, TX}

\author{J. Scott Evans}
\affiliation{Computational Physics Incorporated \\
Springfield, VA}

\author{Michael H. Stevens}
\affiliation{Naval Research Laboratory \\
Washington, DC}

\author{Rebecca N. Schindhelm}
\affiliation{Ball Aerospace \\
Boulder, CO}

\author{Joel Wm. Parker}
\affiliation{Southwest Research Institute \\
1050 Walnut Street, Suite 300 \\
Boulder, CO 80302 , USA}

\author{S. Alan Stern}
\affiliation{Southwest Research Institute \\
San Antonio, TX}

\author{Harold A. Weaver}
\affiliation{Johns Hopkins University Applied Physics Laboratory \\
Columbia, MD}

\author{Catherine B. Olkin}
\affiliation{Southwest Research Institute \\
1050 Walnut Street, Suite 300 \\
Boulder, CO 80302 , USA}

\author{Kimberly Ennico}
\affiliation{SOFIA Science Center \\
Moffett Field, CA}

\author{Jay R. Cummings}
\affiliation{Computational Physics Incorporated \\
Springfield, VA}

\author{G. Randall Gladstone}
\affiliation{Southwest Research Institute \\
San Antonio, TX}

\author{Thomas K. Greathouse}
\affiliation{Southwest Research Institute \\
San Antonio, TX}

\author{David P. Hinson}
\affiliation{SETI Institute \\
Mountain View, CA}

\author{Kurt D. Retherford}
\affiliation{Southwest Research Institute \\
San Antonio, TX}

\author{Michael E. Summers}
\affiliation{George Mason University \\
Fairfax, VA}

\author{Maarten Versteeg}
\affiliation{Southwest Research Institute \\
San Antonio, TX}



\begin{abstract}
  During the New Horizons spacecraft's encounter with Pluto, the Alice
  ultraviolet spectrograph conducted a series of observations that detected
  emissions from both the interplanetary medium (IPM) and Pluto. In the
  direction of Pluto, the IPM was found to be 133.4$\pm$0.6~R at
  Lyman~$\alpha$, 0.24$\pm$0.02~R at Lyman~$\beta$, and $<$0.10~R at
  He\,{\footnotesize~I}~584\AA. We analyzed 3,900~s of data obtained shortly
  before closest approach to Pluto and detect airglow emissions from
  H\,{\footnotesize I}, N\,{\footnotesize~I}, N\,{\footnotesize~II}, N$_2$,
  and CO above the disk of Pluto. We find Pluto's brightness at Lyman-$\alpha$
  to be $29.3\pm1.9$R, in good agreement with pre-encounter estimates. The
  detection of the N\,{\footnotesize~II} multiplet at 1085\AA\ marks the first
  direct detection of ions in Pluto's atmosphere. We do not detect any
  emissions from noble gasses and place a 3$\sigma$ upper limit of 0.14~R on
  the brightness of the Ar\,{\footnotesize~I}~1048\AA\ line. We compare
  pre-encounter model predictions and predictions from our own airglow model,
  based on atmospheric profiles derived from the solar occultation observed by
  New Horizons, to the observed brightness of Pluto's airglow. Although
  completely opaque at Lyman~$\alpha$, Pluto's atmosphere is optically thin at
  wavelengths longer than 1425\AA. Consequently, a significant amount of
  solar FUV light reaches the surface, where it can participate in space
  weathering processes. From the brightness of sunlight reflected from Pluto,
  we find the surface has a reflectance factor (I/F) of 17\% between
  1400-1850\AA. We also report the first detection of an C$_3$ hydrocarbon
  molecule, methylacetylene, in absorption, at a column density of
  $\sim5\times10^{15}$~cm$^{-2}$, corresponding to a column-integrated mixing
  ratio of $1.6\times10^{-6}$.
\end{abstract}



\section{Introduction\label{sec:intro}}

Pluto's atmosphere was definitively discovered in 1988 by the technique of
stellar occultation \citep{Elliotetal89, Hubbardetal88}, but its composition
was unknown. The composition was subsequently deduced to be primarily N$_2$
with trace amounts of CH$_4$ and CO, based on detection of their ices on
Pluto's surface and their vapor pressures \citep{Owenetal93}. Gaseous CH$_4$
was later detected spectroscopically \citep{Youngetal97, Lellouchetal17}. As
of the launch of NASA's New Horizons spacecraft in 2005, only upper limits had
been placed on the amount of atmospheric CO \citep{LYoungetal01,
  Bockelee-Morvanetal01}. The first detections of gaseous CO were claimed in
2011 \citep{Lellouchetal11, Greavesetal11}, and high SNR measurements of
gaseous CO and HCN were made near-in-time to the New Horizons flyby with ALMA
\citep{Lellouchetal17}.

Shortly after the New Horizons spacecraft's closest approach to Pluto, the
Alice instrument observed an occultation of the sun by Pluto
\citep{Gladstoneetal16}, while the Radio Science Experiment (REX) observed an
occultation of Earth \citep{Hinsonetal17}.  From the solar occultation, Alice
detected absorption by N$_2$, CH$_4$, C$_2$H$_2$, C$_2$H$_4$, C$_2$H$_6$, and
haze \citep{Youngetal18}. From the egress of the Earth occultation, which
occurred over the Sputnik Planitia region, REX found that Pluto's atmosphere
was much colder (39~K at the surface, 65-68~K in the upper atmosphere) and
more compact than expected prior to the flyby.

We report here observations of Pluto's atmosphere and surface by the Alice far
ultraviolet (FUV) spectrograph onboard New Horizons,
\citep{Sternetal08NHoverview} just prior to its closest approach.

\section{The Alice FUV Spectrograph}

Alice is a lightweight (4.4kg), low-power (4.4W), imaging, far ultraviolet
(FUV) spectrograph \citep{Sternetal08}.  Sometimes referred to as ``P-Alice''
(for ``PERSI-Alice'', a precursor instrument design, or ``Pluto-Alice''), to
distinguish it from its older sibling instrument on ESA's Rosetta Spacecraft
\citep{Sternetal07ralice} consists an off-axis telescope feeding a 15-cm
diameter Rowland-circle spectrograph with a wavelength range of 520-1870\AA\
and a Nyquist-sampled spectral resolution of 3.8\AA. The detector is an
imaging microchannel plate (MCP) shaped to match the instrument's Rowland
circle, coupled with a double delay line readout anode that converts the
location of the charge cloud produced by the MCP into a 1024x32 (spectral x
spatial) element data array, the central 740x21 pixel region in the data space
maps to the illuminated area of the microchannel plate, with each of the 21
rows subtending a 0.3-degree angle on the sky \citep{Siegmundetal00}. The
front surface of the MCP is coated with a KBr photocathode layer that covers
the passband of roughly 520-1160\AA, a photocathode-free region covering
1160-1280\AA\ and a CsI photocathode region covering 1280-1870\AA. This
photocathode regime was chosen to optimize efficiency at the extremes of the
passband while minimizing the sensitivity to photons from the relatively
bright Lyman~$\alpha$ emission line, which, if not attenuated, would partially
overwhelm the detector electronics.

The Alice instrument has two entrance apertures, the primary 40~mm x 40~mm
square airglow aperture, co-aligned with the LORRI and Ralph instruments, and
a secondary, 1~mm diameter circular solar occultation aperture (SOCC) offset
by $\sim$90\degr\ to the airglow aperture and roughly co-aligned with the
Radio Science Experiment \citep{Tyleretal08} field of view (FOV). The data
analyzed here were all obtained using the airglow aperture.

As seen from the detector, the Alice entrance slit is 6\degr\ long in the
spatial dimension and divided into two sections: a narrow, rectangular region
referred to as the ``stem'' and a wide, square region referred to as the
``box''. The stem is 4\degr\ in angular length along the spatial direction and
0.1\degr\ in the spectral dimension. The optical axis of the instrument is
located 3\degr\ from the bottom of the stem, on the centerline of the
0.1\degr\ -wide slit. The stem portion of the slit corresponds to
approximately rows 6--18 (zero-indexed) in data space, with the center of data
row 16 defining the instrument boresight for spacecraft pointing purposes. The
box is a 2\degr $\times$2\degr\ angular width square, located at the top
(higher detector row numbers) of the stem. The large width of the box was
chosen so that even if there was a significant misalignment between the Alice
solar occultation aperture and REX, both instruments would be able to
simultaneously observe the occultations of the Sun and the Earth by Pluto's
atmosphere. The box portion of the slit corresponds to rows 19-25, with row 18
serving as the transition between the two slit widths.

\section{Airglow Observations and Processing\label{sec:obs}}

\begin{figure*}
\plottwo{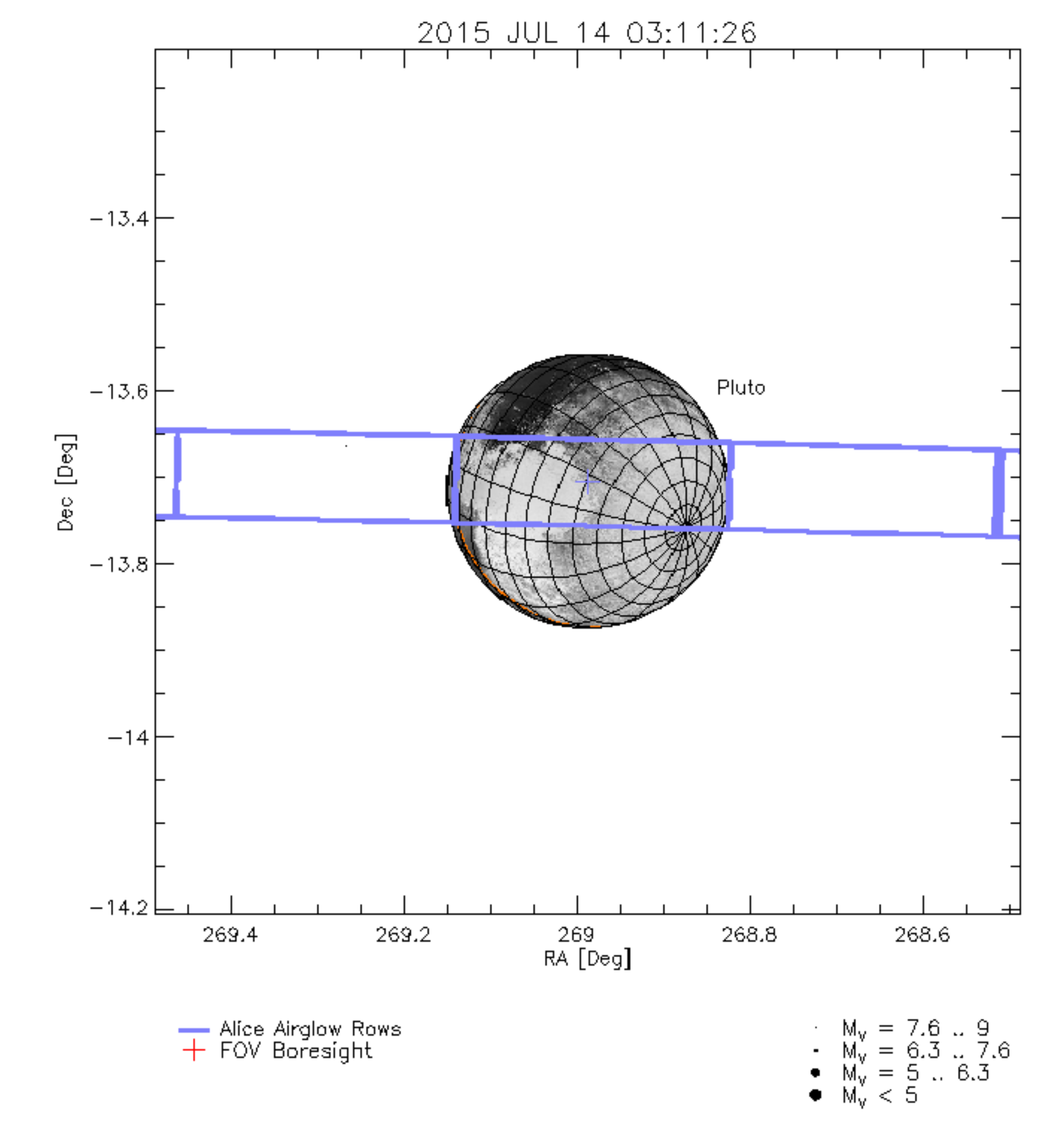}{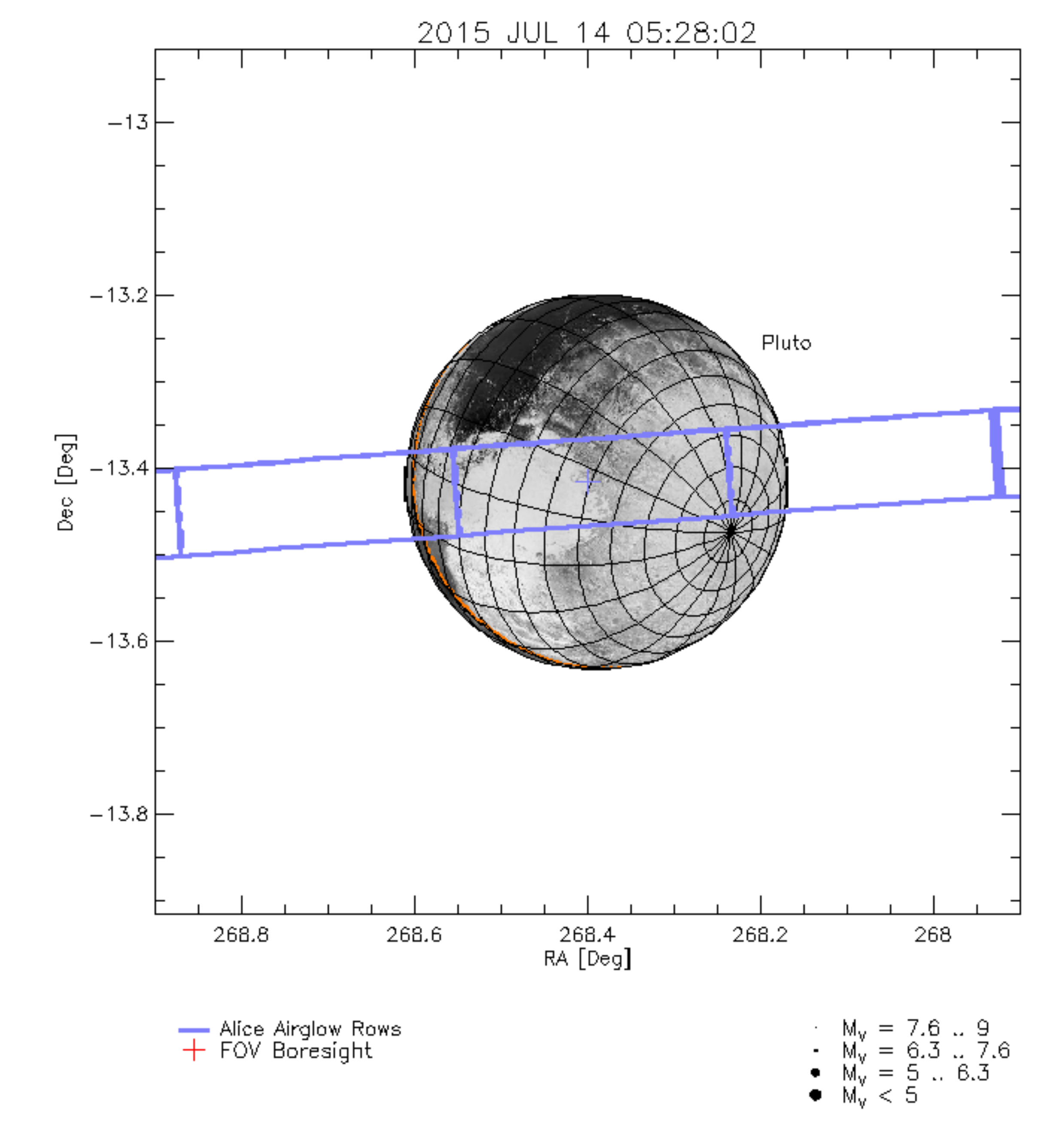}
\caption{Observing geometry during the Airglow3 observation (left) and
  selected exposures from the Airglow4 observation (right). The projection of
  the Alice entrance slit onto the sky plane is outlined in blue. The
  approximately vertical, blue lines delineate the field of view of individual
  detector rows, with row 16 (zero indexed) lying on the disk of Pluto and
  rows 17 and 15 to the left and right of Pluto,
  respectively. \label{fig:geom}}
\end{figure*}

On approach to Pluto, Alice made numerous observations in search of airglow
emissions. For our analysis, we selected data from just two separate airglow
observations, PEAL\_01\_PC\_Airglow\_Appr\_3 and
PEAL\_01\_PC\_Airglow\_Appr\_4, hereafter Airglow3 and Airglow4. These
particular observations were chosen because they are the closest,
long-duration airglow observations of Pluto and thus, presumably, the most
sensitive. Additionally, these observations have a favorable viewing geometry
in which Alice is centered on Pluto, whose disk fills $>$98\% of the field of
view of the central row of the detector (row 16, zero-indexed). this minimizes
the possibility of confusing emission from the interplanetary medium (IPM)
with Plutogenic airglow emissions. The adjacent detector rows, 15 and 17, are
centered at a tangent altitude of $\sim$890 km and cover a region from the
surface to a tangent altitude of 1.6 Pluto radii.  Figure~\ref{fig:geom},
produced using the web-based GeoViz tool \citep{Throopetal09GV}, illustrates
the observing geometry during the Airglow 3 \& 4 observations. The instrument
footprint covers a significant fraction of Sputnik Planitia as well as regions
poleward.

Data from the Airglow3 observation consist of ten 300-second histogram
exposures obtained from 2015 July 14 03:11:26 to 2015 July 14 04:01:26
UTC. During these exposures, the distance to Pluto spanned 427,621--386,306 km
and the phase angle increased slightly from 16.90\degr to 17.09\degr. We also
selected six, 150-second histogram exposures from the 18 exposures of the
Airglow4 observation that covered a similar region on Pluto as Airglow3. These
Airglow4 images were obtained from 2015 July 14 05:20:31 to 2015 July 14
05:35:31 UTC and spanned 320,976--308,588 km in distance and 17.51-17.60\degr\
in phase angle. To maximize the signal-to-noise ratio in the data, we co-added
all 16 spectra. Observational details are shown in Table~\ref{tab:obs}.

\begin{deluxetable*}{lcccc}
 
 \tablecaption{Details of Airglow Observations \label{tab:obs}}
 \tablecolumns{3}
  \tablewidth{0pt}
  \tablehead{
    \colhead{Observational Quantity} &
    \colhead{Airglow3} &
    \colhead{Airglow4}
  }
  \startdata
  Number of integrations & 10 & 6 \\
  Total integration time (s) & 3,000 & 900 \\
  Start Time  & 2015 July 14 03:11:26 &  2015 July 14 05:20:31 \\
  End Time & 2015 July 14 04:01:26 & 2015 July 14 05:35:31 \\
  Distance to Pluto at start (km) & 427,600 & 386,300 \\
  Distance to Pluto at end (km) & 321,000 & 308,600 \\
  Pluto phase angle at start (\degr) & 16.90 & 17.09 \\
  Pluto phase angle at end (\degr) & 17.51 & 17.60 \\
  \enddata
\end{deluxetable*}

We apply the standard Alice data reduction techniques of dead time correction,
stim pixel correction, and dark subtraction. These are described in more
detail in the Appendix, below. As discussed above, there is no photocathode
coating on the microchannel plate in the region around Lyman~$\alpha$
(1216\AA). This causes the extended wings of the line profile to appear to
drop to zero, increase sharply around 60~\AA\ from Lyman~$\alpha$ (where the
CsI and KBr photocathode coatings begin) and then decrease gradually with
further distance from the line center. Because Lyman~$\alpha$ emission line is
so intrinsically bright, the extended wings of the line profile are comparable
in intensity to the faint airglow emissions we are searching for--even several
hundred angstroms away from core of the line. Thus, careful removal of the
scattered Lyman~$\alpha$ profile is required. 

We created a Lyman~$\alpha$ template image by summing 38 hours of Alice
observations (PC\_AIRGLOW\_DOY, where DOY is the day of year), made on
approach to Pluto between 2015 May 29 (DOY 149) and 2015 June 18 (DOY
169). Owing to the large distance of New Horizons to Pluto (greater than 30
million kilometers), no airglow emissions or sunlight reflected from Pluto
were detected in these data, and Pluto's disk blocks out an insignificant
fraction of the field of view. In half of these observations, Pluto was placed
at the center of the box portion of the slit and in the other half, Pluto was
placed at the instrument boresight in the stem. No significant differences
were seen between the two pointings. IPM emission lines were detected at
Lyman~$\alpha$ (1216\AA) and Lyman~$\beta$ (1026\AA) but not at
He\,{\footnotesize I}~584\AA. The observed brightness (or upper limit) of
these lines is given in Table~\ref{tab:ipm}. The brightness of the IPM
Lyman~$\alpha$ is $\sim$1.5$\times$ brighter than pre-encounter predictions
\citep{Gladstoneetal15}.

\begin{deluxetable*}{lcc}[b]
  \tablecaption{Brightness of IPM lines from New Horizons (r=32.6~AU) in the
    direction of Pluto\tablenotemark{a} \label{tab:ipm}}
  \tablecolumns{3}
  \tablewidth{0pt}
  \tablehead{
    \colhead{Species} &
    \colhead{Wavelength (\AA)} &
    \colhead{Observed Intensity\tablenotemark{b} (R)}
  }
  \startdata
  He\,{\footnotesize I} & 584 & $<$ 0.10 \\
  H\,{\footnotesize I} & 1026 & 0.24 $\pm$ 0.02 \\
  H\,{\footnotesize I} & 1216 & 133.4 $\pm$ 0.6 \\
  \enddata
  \tablenotetext{a}{$\alpha$=18$^h$2$^m$38.7$^s$, $\delta$ = -14\degr
    37\arcmin37\farcs2}
  \tablenotetext{b}{Quoted error bars are 1$\sigma$, while
    the He\,{\footnotesize I}~584\AA\ upper limit is 3$\sigma$}
\end{deluxetable*}

\begin{figure*}
\plotone{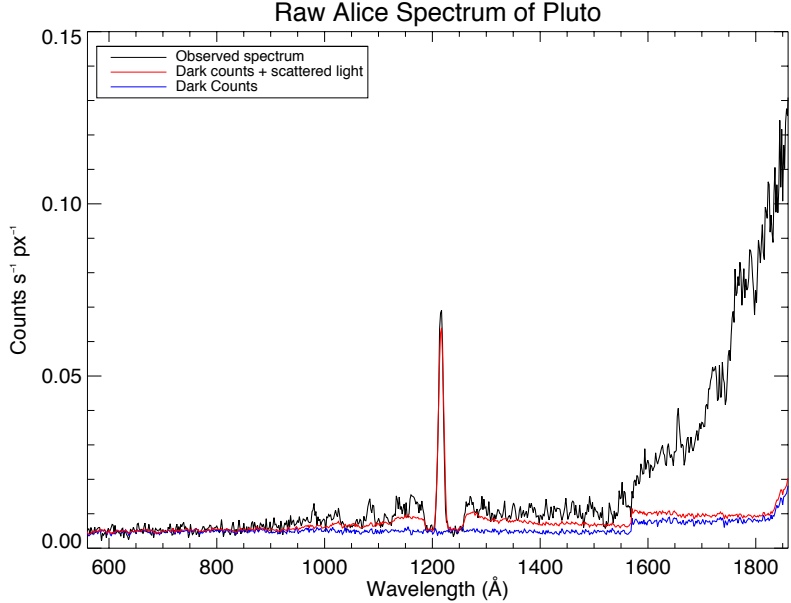}
\caption{Alice count rate spectra of Pluto and background sources. The spectra
  are extracted from detector row 16, except for wavelengths longer than
  1570\AA, which also include row 15. This causes the apparent jump in
  background rates. The black curve shows the uncalibrated count rate spectrum
  from the sum of the Airglow3 and Airglow4 observations used in this
  analysis. The blue curve shows the dark count background produced by the
  detector when the airglow aperture door is closed, and the red curve shows
  the sum of the dark count spectrum and the IPM Lyman~$\alpha$ profile,
  scaled to the level of Lyman~$\alpha$ in the Airglow3
  observations.\label{fig:raw}}
\end{figure*}

Since interplanetary Lyman~$\beta$ emission is $\sim$500x fainter than
Lyman~$\alpha$ emission, the extended wings of its line profile are not
significant. To prevent the unintentional subtraction of the IPM Lyman~$\beta$
signal from the Pluto observations, we fit a Gaussian line profile with a
linear background, to the IPM spectrum around 1026\AA\ and subtract off the
Gaussian component. The resulting template image contains only interplanetary
Lyman~$\alpha$ and detector dark counts. We remove these dark counts by
subtracting a composite ``dark'' image, obtained while the airglow aperture
door was closed. The spectrum of these dark counts can be seen in the blue
line of Figure~\ref{fig:raw}. After subtracting the scaled dark image from the
Pluto observations, we normalize the IPM Lyman~$\alpha$ template to the
brightness of the Lyman~$\alpha$ emission in the Pluto data.

Due to the slight misalignment of the Alice detector and the optical axes of
the spectrograph, emissions that are centered in a given detector row at short
wavelengths will partially spill over onto the next lower detector row for
wavelengths greater than $\sim$1570\AA. We therefore extract the airglow
spectrum from row 16 and add the spectrum of row 15 to it for
$\lambda > 1570$\AA. The resulting uncalibrated spectrum of Pluto is shown in
Figure~\ref{fig:raw}.  We divide by the instrument's effective area curve and
by the solid angle of the sky as seen by a single detector row (0.3$\degr$
spatial along the slit by 0.1$\degr$ across it) to calibrate the spectrum in
units of radiance.

\section{Pluto's Atmospheric Transmission and Surface
  Reflectance \label{sec:surf}}

\begin{figure*}
\plotone{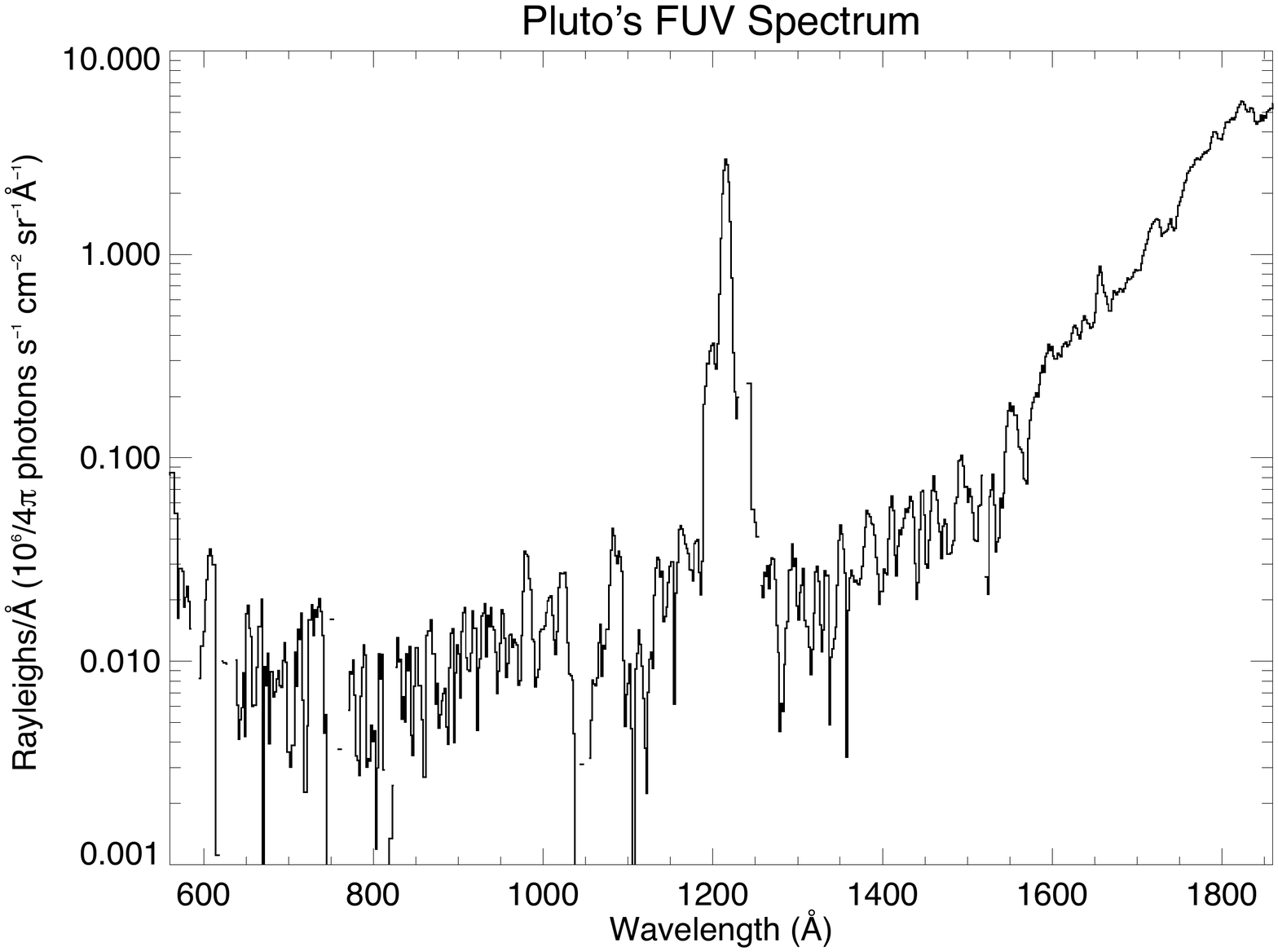}
\caption{Observed spectral radiance of Pluto over the Alice passband. Both
  dark counts and a scattered Lyman~$\alpha$ profile have been subtracted from
  the data. The dramatic increase in flux at the long wavelength end of the
  spectrum is the result of sunlight reflected from the surface of Pluto with
  a small contribution from sunlight scattered by atmospheric haze
  particles.  \label{fig:logspec}}
\end{figure*}

The observed spectral radiance of Pluto over the Alice passband is shown in
Figure~\ref{fig:logspec}. The central spectral feature is the Lyman~$\alpha$
airglow at 1216\AA. Since the disk of Pluto completely fills the FOV of the
central row during this observation, there is no contribution from the
IPM. Faint airglow emission features arising from molecular, atomic and
ionized species (e.g., N$_2$~CY(0,1)~980\AA, N\,{\footnotesize I}~1493\AA, and
N\,{\footnotesize II}~1085\AA) are present in the spectrum and are discussed
in Section~\ref{sec:airglow}, below.

At wavelengths greater than $\sim$1500\AA, Pluto's spectrum is dominated by
sunlight passing through the atmosphere and reflecting off the
surface. Scattering of sunlight by atmospheric haze particles 100-200~km above
the surface may also contribute a small amount to the observed spectrum at
these wavelengths. For our purposes, the reflected/scattered solar light
serves as an additional source of background that potentially obscures fainter
airglow emissions in this region of the spectrum. We therefore attempt to
remove the solar contribution by constructing a simple model of Pluto's
atmospheric transmission and surface reflectance, based on the solar
occultation profiles of Pluto's atmosphere \citep{Youngetal18}.

\subsection{Atmospheric Transmission \label{sec:trans}}

\begin{figure*}
\plotone{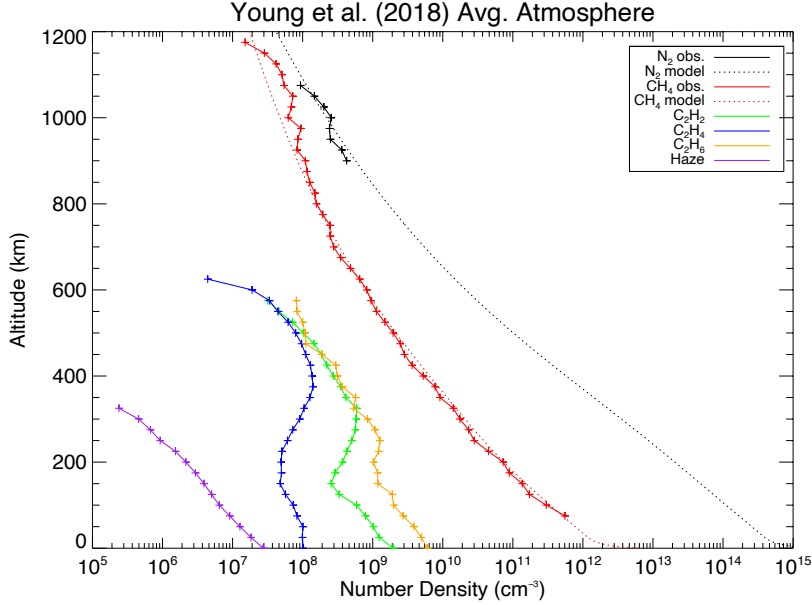}
\caption{Atmospheric profiles of Pluto based on the solar occultation
  observations of \citet{Youngetal18}. Dotted lines represent model fits to
  the observed N$_2$ and CH$_4$ profiles. Haze density is derived by assuming
  a wavelength-independent cross section of $1\times10^{-15}$
  cm$^2$. \label{fig:atmprofile}}
\end{figure*}

The ingress and egress solar occultation profiles reported by
\citet{Youngetal18} are largely similar. To increase the signal-to-noise in
our analysis, we averaged the two occultation profiles together and re-binned
the data to 25~km vertical resolution. We used this profile to reproduce the
results of \citet{Youngetal18}. Pluto's hazes are complex
\citep{Changetal17,Zhangetal17, Krasnopolsky20}, and we do not attempt to
derive a column density directly from the occultation profiles. Rather,
following \citet{Youngetal18}, we treat the haze as a spectrally-neutral
source of atmospheric opacity in our fits to the solar occultation
transmission spectrum. For comparison with other atmospheric constituents, we
assume a wavelength-independent haze cross section of
1$\times$10$^{-15}$~cm$^2$.  With the additional assumption of spherical
symmetry (reasonable, given the similarity of the ingress and egress profiles)
we can apply the Abel transform \citep{Roble:Hays72} to derive local number
density from line-of-sight abundances (column densities):

\begin{equation}\label{eq:abel}
  n(r) = - \frac{1}{\pi} \int_r^\infty \frac{\left[dN(r')/dr'\right]}{\sqrt{r'^2-r^2}}dr' ,
\end{equation}

\noindent where $n(r)$ is the number density of a given species at radial
distance, $r$, from Pluto and $N(r')$ is the column density of that species at
a tangent radius, $r'$. The resulting atmospheric profiles are shown in
Figure~\ref{fig:atmprofile}, which is broadly similar to Figure~17 of
\citet{Youngetal18}.

\begin{figure*}
\plotone{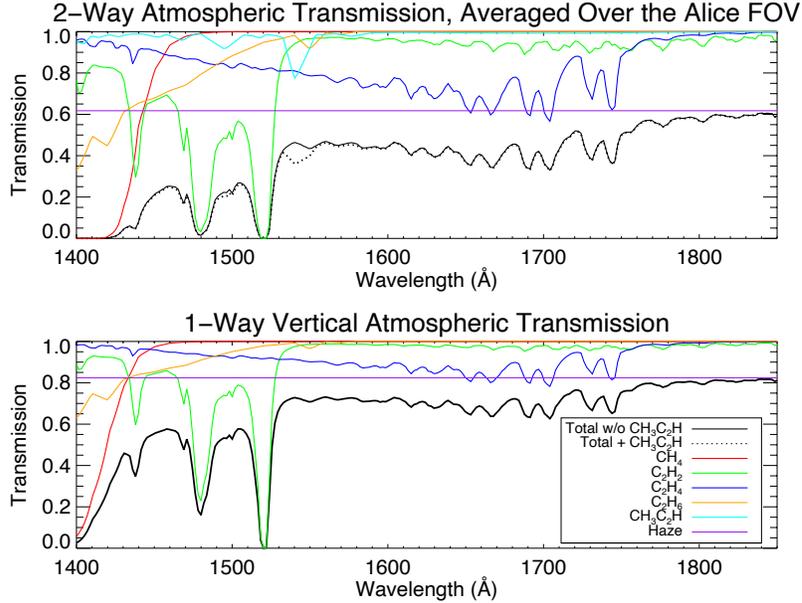}
\caption{Top: 2-way transmission through Pluto’s atmosphere, averaged over the
  Alice field of view. We include atmospheric absorption from methane
  (CH$_4$), acetylene (C$_2$H$_2$), ethylene (C$_2$H$_4$), ethane
  (C$_2$H$_6$), and haze particles. The top panel also shows the additional
  absorption by $\sim5\times10^{15}$~cm$^{-2}$ of methylacetylene (C$_3$H$_4$
  along the line-of-sight (see discussion in Section
  \ref{sec:methylacetylene}). The detailed properties of the haze are not
  modeled. Instead, the haze is assumed to have a wavelength-independent cross
  section of $1\times10^{-15}$ cm$^2$ for rough comparisons to other
  atmospheric constituents. Bottom: 1-way vertical atmospheric
  transmission. \label{fig:trans}}
\end{figure*}

As shown in Figure~\ref{fig:geom}, the Alice field of view covered a
significant fraction of the disk of Pluto during the Airglow 3 and 4
observations, and as such covers a large range of solar incidence and emission
angles. Therefore, we divided the field of view of the central detector row
(row 16, zero-indexed) into a grid of 31x11 lines-of-sight, separated by
0.01$\degr$ (corresponding to a separation of roughly 70~km on Pluto's
surface), and calculated the atmospheric absorption and surface reflection
along each line of sight. Averaged over the Alice field-of-view, the solar
incidence angle is 36.2$\degr$, while the average emission angle is
33.3$\degr$. We integrate the atmospheric profiles shown in
Fig.~\ref{fig:atmprofile} along the path from the sun to the surface and then
from the surface to Alice for each of the 341 lines-of-sight. The two-way
atmospheric transmission, averaged over the Alice field of view, is shown in
the top panel of Figure~\ref{fig:trans}, while the one-way vertical
atmospheric transmission is shown in the bottom panel (cf. Figure~12 of
\citet{Youngetal18}).

Pluto's atmosphere is completely opaque at wavelengths below 1400\AA, largely
due to absorption by methane (CH$_4$). However, the absorption cross-section
of methane decreases by more than four orders of magnitude between 1400~\AA\
and 1500~\AA, which results in the vertical optical depth of the atmosphere,
$\tau_{v}$, being less than one for $\lambda > 1425$~\AA. Acetylene
(C$_2$H$_2$) is the primary atmospheric absorber between 1430-1530~\AA, while
at wavelengths greater than 1530~\AA, the atmospheric transmission is
controlled by both haze particles and ethylene (C$_2$H$_4$).

As shown in the bottom panel of Figure~\ref{fig:trans} photons with
$\lambda > 1425$\AA\ readily pass through the atmosphere and interact with
Pluto's surface (i.e. $\tau_v < 1$). These photons have only $\sim$25\% less
energy than Lyman~$\alpha$ photons--enough to break molecular bonds and drive
photolysis. \citep{Olkinetal17} report that Pluto's equatorial regions, which
receive greater insolation when averaged over Pluto's orbit, are darker and
redder than the poles, which are brighter and more neutral in color. They
propose that this surface color distribution could be produced by the
transport of volatiles away from the warmer equator towards the colder
poles. We suggest that in addition to this mechanism, longer-wavelength FUV
photons photolyze tholins and haze particles on the surface, further reddening
the equatorial regions.

\subsection{Pluto's FUV Surface Reflectance \label{sec:reflectance}}

The reflectance factor, sometimes referred to as I/F, is defined as:

\begin{equation}\label{eq:ioverf}
I/F (\lambda) = \frac{\pi h c I (\lambda,\phi) r^2}{\mu_0 \Omega F_\sun
  (\lambda)}
\end{equation}

\noindent where $h$ is Planck's constant; $c$ is the speed of light; $I$ is
the observed spectral intensity in photons~s$^{-1}$~cm$^{-2}$~\AA$^{-1}$ as a
function of wavelength, $\lambda$, at a given solar phase angle, $\phi$; $r$
is the heliocentric distance of Pluto in AU; $\mu_0$ is the cosine of the
solar incidence angle, averaged over the field of view;
$\Omega=9.1\times10^{-6}$ Sr is the solid angle subtended by a single row of
the Alice detector; and $F_\sun$ is the solar flux at 1~AU. After proper
calibration, $I(\lambda,\phi)$ is what is actually measured by Alice. We use
the same $F_\sun(\lambda)$ as in \citet{Youngetal18}, which was assembled from
SUMER reference spectra \citep{Curdtetal01} and observations from TIMED/SEE
\citep{Woodsetal05}.

\begin{figure*}
\plotone{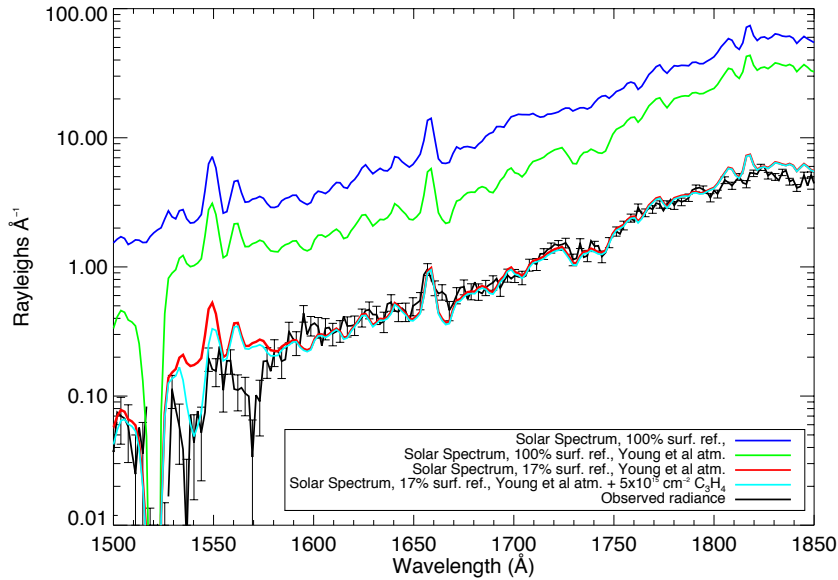}
\caption{The observed spectral radiance of Pluto and solar spectra with
  various combinations of surface reflectance and atmospheric absorption. The
  spectral radiance observed by Alice is shown in black, with 1$\sigma$ error
  bars shown for every other point. If Pluto had no atmosphere and a surface
  reflectance of 100\%, Alice would observe the blue curve. Including
  absorption by the \citet{Youngetal18} atmosphere yields the green curve,
  while also reducing the surface reflectance to 17\% produces the red
  curve--a surprisingly good match to the data for wavelengths longer than
  1580\AA.  Between 1530-1580\AA, there is an additional source/sources of
  absorption of the solar spectrum not included in \citet{Youngetal18} that
  results in an over-subtraction of the solar spectrum. Including absorption
  by methylacetylene (C$_3$H$_4$ or propyne) with a 2-way column density of
  $\sim5\times10^{15}$~cm$^{-2}$ significantly improves the model fit between
  1535--1550\AA.  \label{fig:fuvref}}
\end{figure*}

In the absence of an atmosphere and with a surface reflectance factor of
100\%, the spectral radiance of Pluto would be about an order of magnitude
greater than what is observed, as shown with the blue curve of
Figure~\ref{fig:fuvref}. Including the 2-way atmospheric transmission shown in
Fig.~\ref{fig:trans} reduces our model spectral radiance to about 6$\times$
what is observed at long wavelengths. Combining this with a
wavelength-independent surface reflectance factor of 17\% yields a
surprisingly good match to the observed radiance of Pluto at wavelengths
greater than $\sim$1570\AA. Notably, the feature at 1657\AA\ appears to be
entirely due to the reflected/scattered solar C\,{\footnotesize I} multiplet
and not airglow emission from Pluto's atmosphere.

Compared to other planetary surfaces, an FUV I/F of 17\% is relatively
high. For example, comet 67P/Churyumov-Gerasimenko has an I/F of just 1-2\%
\citep{Sternetal15ralice}; Saturn's moon, Phoebe, has a reflectance between
1-3\% \citep{Hendrix:hansen08}; while the Moon's I/F varies between 2-10\%
\citep{Gladstoneetal12}.

As discussed above, Pluto's atmospheric haze was treated simply as a source of
extinction. In reality, haze particles will both scatter and absorb
sunlight. Therefore, the 17\% surface I/F value should really be thought of as
an upper limit. Modeling the properties of Pluto's atmospheric haze is well
beyond the scope of this paper. However, as an end-member case, if we assume
the haze particles are simple 0.2$\micron$ spheres with a single scattering
albedo of 0.55 at 1500\AA, Mie scattering theory predicts that the haze will
be roughly half absorbing and half scattering and that the scattering should
be roughly independent with wavelength over the Alice pass band. If we then
assume that as much sunlight is scattered back into our line-of-sight as out
of it, a surface I/F value of 13\% is required to match the Alice
observations. We suggest that a more detailed analysis of Pluto's ultraviolet
surface reflectance, properly accounting for atmospheric haze, is a fruitful
area for subsequent work.

\subsection{Methylacetylene \label{sec:methylacetylene}}

While the overall match between the Alice observations and our simple
transmission/reflectance model is fairly good, as shown in
Fig.~\ref{fig:fuvref}, our simple model predicts significantly more flux than
is observed between 1535--1570\AA, resulting in an over-subtraction of the
solar spectrum. This suggests that either the atmosphere or surface has one or
more additional sources of opacity/absorption. However, FUV absorption
features from solids tend to be very broad, on the order of 100's of \AA
ngstroms \citep{Wagneretal87}. Likewise, we are not aware of any mechanism by
which atmospheric haze can produce such relatively narrow absorption
features. Thus, we favor the interpretation that one or more additional
gaseous species are present in Pluto's atmosphere at high-enough column
densities to significantly absorb sunlight passing through the atmosphere.

We examined the absorption cross sections of 32 additional atomic and
molecular species that might plausibly be found in Pluto's atmosphere (C,
CH$_3$, CH$_4$O, C$_2$H, C$_2$H$_5$, C$_3$H$_3$, C$_3$H$_4$ (both the allene
and methylacetylene isomers), C$_3$H$_6$ (both the propene and cyclopropane
isomers), C$_3$H$_8$, C$_4$H$_2$, C$_4$H$_4$, C$_6$H$_6$, CO, CO$_2$, H,
H$_2$, H$_2$CO, H$_2$O, H$_2$O$_2$, HCN, HC$_3$N, HNCO, N, NH$_3$, O, O$_2$,
O$_3$, OCS, PH$_3$, SO, and SO$_2$), and found that, among them, only
methylacetylene (\mbox{H$_3$C$-$C$\equiv$CH}, or propyne), has both a strong
absorption band in this region and a lack of strong absorption bands at longer
wavelengths, where no additional absorption is seen. Methylacetylene has been
observed in the upper atmospheres of both Titan, where it can reach local
mixing ratios of up to 10$^{-5}$ \citep{Lietal15}, and Jupiter, where it is
part of an important chemical pathway in the production of acetylene
(C$_2$H$_2$) \citep{Gladstoneetal96}. The Pluto photochemical model of
\citet{Wongetal17} predicts a methylacetylene column density of
$2.5\times10^{15}$~cm$^2$ along our two-path line-of-sight. Including
methylacetylene in our model at twice this value ($5\times10^{15}$~cm$^{-2}$;
corresponding to a column-integrated mixing ratio of $1.6\times10^{-6}$)
produces a significantly better fit around 1540\AA, as shown in
Figure~\ref{fig:fuvref}.

If absorption of reflected sunlight by methylacetylene is detectable in the
airglow observations, it should also be evident in the solar occultation
observed by Alice. Careful re-examination of the solar occultation profiles
described by \citet{Youngetal18} over a range of tangent altitudes from
100-150~km shows a previously-unnoticed absorption feature at 1540\AA,
consistent with methylacetylene. Our preliminary re-analysis yields a
methylacetylene column density of $1.5\times10^{15}$~cm$^{-2}$ along this line
of sight. Since the analysis of the solar occultation observations involves
only the transmission of sunlight through Pluto's atmosphere and we directly
measure the unocculted solar spectrum, we can exclude the possibilities that
this feature is caused by something in Pluto's surface reflectance spectrum or
the solar spectrum itself. We therefore claim the first detection of a
C$_3$-hydrocarbon in Pluto's atmosphere and suggest that an additional,
yet-unidentified, atmospheric species is responsible for the apparent
absorption features at 1530\AA\ and 1570\AA. The level of methylacetylene
should provide an important constraint for future photochemical models of
Pluto's atmosphere. 

\section{Pluto's Airglow Emissions \label{sec:airglow}}

\begin{figure*}
\plotone{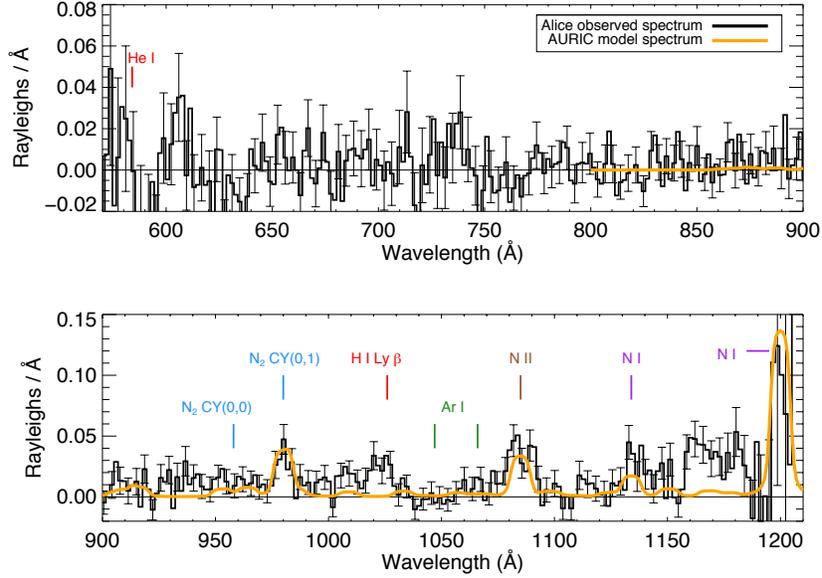}
\caption{EUV airglow spectrum of Pluto.  For clarity, every other error bar is
  plotted, representing the 1$\sigma$ statistical uncertainty. At these
  wavelengths, Pluto’s atmosphere is opaque, and no reflected sunlight is
  present in the spectrum.  Emissions from N$_2$, N\,{\footnotesize II},
  N\,{\footnotesize I}, and H\,{\footnotesize I}, are clearly
  detected. Notably absent are emission lines from argon at 1048\AA\ and
  1067\AA. The orange curve is a synthetic spectrum produced by our AURIC
  model using the atmospheric profiles of \citet{Youngetal18} combined with
  surface mixing ratios of 5.0$\times$10$^{-4}$ for CO and
  1.5$\times$10$^{-4}$ for Ar\,{\footnotesize I}. Our model does not include
  emissions from hydrogen. The elevated brightness between 1160--1190\AA\ is
  likely an artifact of the Lyman-$\alpha$ background
  subtraction. \label{fig:euv_spec}}
\end{figure*}

\begin{figure*}
\plotone{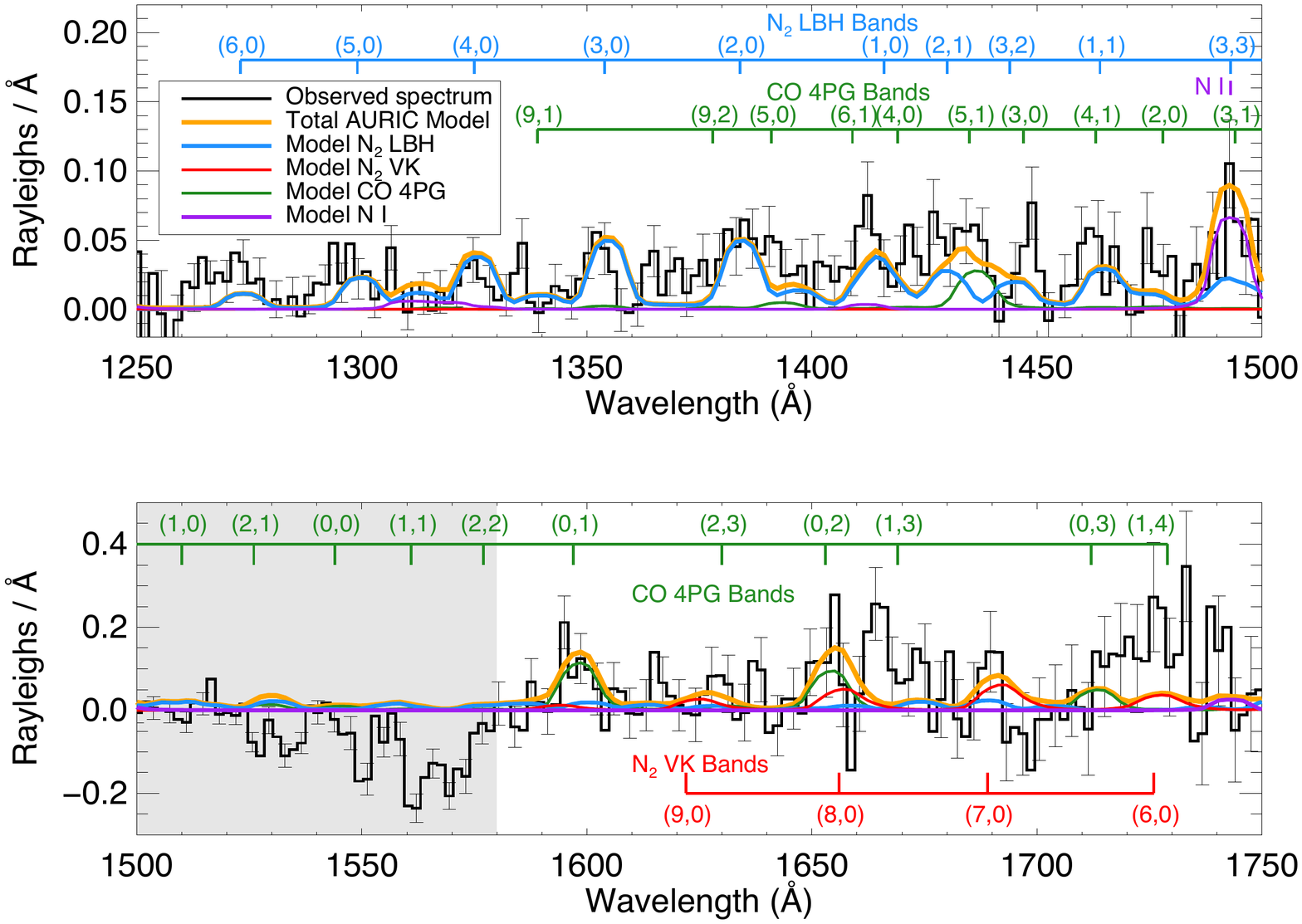}
\caption{FUV airglow spectrum of Pluto with our AURIC model prediction.  For
  clarity, every other error bar is plotted, representing the 1$\sigma$
  statistical uncertainty. Emissions from N$_2$, N\,{\footnotesize I} and CO
  are detected.  The orange curve is a synthetic spectrum produced by our
  AURIC model using the atmospheric profiles of \citet{Youngetal18} combined
  with surface mixing ratios of 5.0$\times$10$^{-4}$ for CO and
  1.5$\times$10$^{-4}$ for Ar\,{\footnotesize I}. Between 1500--1580\AA, there
  is a source of additional absorption in Pluto's atmosphere or surface that
  is not included in our background model, resulting in an over-subtraction of
  the reflected solar spectrum (see Sec.~\ref{sec:surf}). This region, shaded
  in gray, is excluded from further analysis. \label{fig:fuv_spec}}
\end{figure*}

Pluto's extreme ultraviolet (EUV) airglow spectrum is shown in
Figure~\ref{fig:euv_spec} and its far ultraviolet (FUV) spectrum, after
subtracting the reflected solar spectrum, is shown in
Figure~\ref{fig:fuv_spec}. (We loosely define the EUV region as wavelengths
shorter than Lyman-$\alpha$ at 1216\AA\ and the FUV as longer than
Lyman-$\alpha$ but shorter than 2000\AA.) Although the signal-to-noise ratio
of the spectrum is relatively low throughout much of the bandpass, (observed
count rates are generally on the order of 1 count/pixel/100~seconds), faint
emission features from H\,{\footnotesize I}, N\,{\footnotesize II},
N\,{\footnotesize I}, N$_2$, and CO are detected at brightnesses of a few
tenths of a Rayleigh
($\frac{10^6}{4\pi}$~photons~s$^{-1}$~cm$^{-2}$~sr$^{-1}$). We discuss
individual species in further detail in the subsections below.

To aid the identification of these features, we compare the observed airglow
spectrum to a model spectrum produced by a version of the Atmospheric
Ultraviolet Radiance Integrated Code (AURIC) \citep{Stricklandetal99,
  Stevensetal11, Stevensetal15, Evansetal15} adapted to Pluto. AURIC generates
emission spectra from multiple species as a function of viewing direction. In
particular, it calculates emissions from solar fluorescence, electron impact,
photoionization, photodissociation, and recombination and then propagates
these emissions through a radiative transfer model of the atmosphere of
interest. We generate emission spectra over the 800-2000\AA\ bandpass. Since
the Alice slit covers a significant fraction of the disk of Pluto, as
discussed in Sec.~\ref{sec:trans}, we average the model output over an
evenly-spaced grid of 341 lines of sight, separated by 0.01\degr\ ($\sim$70~km
projected on the surface). We start with the atmosphere of
\citet{Youngetal18}, as shown in Fig.~\ref{fig:atmprofile} and add CO at a
surface mixing ratio of 5.0$\times$10$^{-4}$ \citep{Lellouchetal17}, assuming
that, like the rest of the atmosphere, it is in gravitational diffusive
equilibrium.

AURIC produces synthetic spectra for molecular emission band systems and
individual atomic species. For this paper, we did not vary the model's input
parameters in an attempt to fit the observed airglow spectrum in a fully
self-consistent manner. Such work will be the subject of a future
publication. However, recent analyses of the atmospheres of Mars and Titan
using AURIC \citep{Stevensetal15, Jainetal15, Schneideretal15, Stevensetal17},
have shown that weighting each of the individual component spectra can result
in a markedly better match to the data. We use a multiple linear regression
(MLR) algorithm to determine the weights for each component spectrum:

\begin{equation}\label{eq:mlr}
  S_{\mbox{final}}(\lambda) = \sum\limits^n a_1 S_1(\lambda) + a_2
  S_2(\lambda) + \cdots + a_n S_n(\lambda)
\end{equation}

\noindent where $a_n$ are the weights for the individual component spectra,
$S_n(\lambda)$, for each molecular band system or atom produced by our AURIC
model. Our MLR fit has five component spectra: the Lyman-Birge-Hopfield (LBH)
and Vegard-Kaplan (VK) band systems of N$_2$, the Fourth Positive (4PG) and
Hopfield-Birge (HB) band systems of CO, and the emission multiplets of
N~{\footnotesize I}. The weights were simultaneously fit to the data over the
following three bandpasses: 1100\AA~$< \lambda <$~1200\AA,
1270\AA~$ < \lambda <$~1505\AA, and 1580\AA~$<\lambda<$~1750\AA. The model
intensities of these emissions, integrated over 800\AA~$< \lambda <$~2000\AA\
are given in Table~\ref{tab:mlr}.

\begin{deluxetable*}{lcc}
  \small
\tablecaption{Integrated MLR model brightnesses (800\AA\ $< \lambda <$
  2000\AA) \label{tab:mlr}}
\tablecolumns{3}
\tablewidth{0pt}
\tablehead{
  \colhead{Emission} &
  \colhead{Intensity (R)} &
  \colhead{1$\sigma$ error (R)}
}
\startdata
N$_2$ LBH bands & 7.8 &  0.7 \\
N$_2$ VK bands & 3.3 & 1.6  \\
CO 4PG bands & 4.1 & 0.9 \\
CO HB bands & 0.17 & 0.08 \\
N~{\footnotesize I} &  4.3 & 0.7 \\
\enddata
\end{deluxetable*}

Our model spectrum is a reasonably good match to the observations, as can be
seen in Figures~\ref{fig:euv_spec} and \ref{fig:fuv_spec}. All of the
detectable emission features predicted by our model are either present in, or
at least consistent with, the observed airglow spectrum. Conversely, there are
no significant emission features predicted by our model that are clearly
missing from the observed spectrum. On the other hand, there are several
features in the observed spectrum that appear to be statistically significant
but do not appear in our model. These are are likely either instrumental
artifacts or systematic effects of our data processing, and they are discussed
in further detail in Section~\ref{sec:leftovers}, below.

From the results of our AURIC model, we claim the detection ($>4\sigma$
likelihood) of the N$_2$~LBH and CO~4PG bands. With a 2$\sigma$ level of
confidence, we also plausibly detect emissions from the N$_2$~VK bands, the
brightest of which is predicted to be the (7,0) band at 1689\AA. However, we
advise caution in interpreting the CO~HB bands as a ($\sim$2$\sigma$)
``detection''. This band system in the EUV is quite faint ($<$0.2~R,
integrated over the entire instrument bandpass), and while including it in the
model does result in a statistically better fit, none of the predicted CO~HB
emissions (the brightest of which occur at 1151\AA\ and 1124\AA) are
particularly compelling. We attempted to include the Birge-Hopfield-1 band
system of N$_2$ and Cameron band system of CO in our MLR model, but found that
these band systems were not sufficiently constrained by our data.

We detect airlow emissions from H\,{\footnotesize I} at both Lyman~$\alpha$
and Lyman~$\beta$ and derive the amount of hydrogen above the $\tau=1$
altitude in Section~\ref{sec:H}, below. Although this is an important
constraint, we cannot measure the vertical profile of hydrogen in Pluto's
atmosphere from the Alice data or from any other currently existing
observations. Since hydrogen is produced and destroyed by a large number of
reactions in Pluto's atmosphere and developing a full photochemical model is
well beyond the scope of this observational paper, we do not include it in our
AURIC model atmosphere. As neither H nor H$_2$ are a significant source of
atmospheric opacity --at least for any physically reasonable amount of
hydrogen--this omission does not affect the interpretation of other airglow
spectral features.

In the EUV region of the spectrum, the airglow emission features are
well-separated (see Fig~\ref{fig:euv_spec}). To determine their brightness (or
upper limits), we fit a Gaussian line profile plus a linear background to each
feature. The width of the Gaussian profile was held constant to match the line
function of the instrument for a filled slit \citep {Sternetal08} . The area
under the fitted Gaussian profile (or the 3$\sigma$ uncertainty thereof, in
the case of a non-detection) is given in Table~\ref{tab:emissions}, along with
model predictions from \citet{Summersetal97}, \citet{Sternetal08},
\citet{Youngetal08}, \citet{Stevensetal13}, \citet{Jain:Bhardwaj15}, and our
MLR-weighted AURIC model.

\begin{deluxetable*}{lcccccccc}[b]
  \tablecaption{Brightness of selected airglow emission features\label{tab:emissions}}
  \small
  \tablecolumns{9}
  \tablewidth{0pt}
  \tablehead{
    \colhead{Species} &
    \colhead{Wavelength} &
    \colhead{Intensity\tablenotemark{a}} &
    \multicolumn{6}{c}{Model Predictions (R)} \\
    \colhead{} &
    \colhead{(\AA)} &
    \colhead{(R)} & 
    \colhead{SSG97\tablenotemark{b}} &
    \colhead{SSS08\tablenotemark{c}} &
    \colhead{YSW08\tablenotemark{d}} & 
    \colhead{SEG13\tablenotemark{e}} &
    \colhead{JB15\tablenotemark{f}} & 
    \colhead{this work}
  }
  \startdata
  He\,{\footnotesize I} & \, 584 & $<$ 0.49 & -- & -- & -- & -- & -- & -- \\
  N\,{\footnotesize II} & \, 916 & $<$ 0.21 & 0.08  & 0.13 & 0.04 & 0.05 & -- & 0.08 \\
  N$_2$ CY(0,0) & \, 958 & $<$ 0.20 & 0.7  & 1.3 & 0.35 & 0.0 & -- & 0.0\\
  N$_2$ CY(0,1) & \, 980 & 0.28 $\pm$ 0.08 & -- & -- & -- & 0.2 & -- & 0.4\\
  H\,{\footnotesize I} & 1026 & 0.20 $\pm$ 0.04 & -- & -- & -- & -- & -- & -- \\
  Ar\,{\footnotesize I} & 1048 &  $<$ 0.14 & 0.3  & 0.45 & 0.15 & 0.3 & -- & $2\times10^{-4}$ \\
  Ar\,{\footnotesize I} & 1067 &  $<$ 0.21 & 0.3  & 0.35 & 0.15 & 0.3 & -- & $2\times10^{-4}$ \\
  N\,{\footnotesize II} & 1085 & 0.57 $\pm$ 0.14 & 0.4 & 0.6 & 0.2 & 0.2 & -- & 0.30 \\
  N\,{\footnotesize I} & 1134 & 0.25 $\pm$ 0.09 & 0.2 & 0.9 & 0.1 & 0.1 & -- & 0.15 \\
  N\,{\footnotesize I} & 1200 & 0.66 $\pm$ 0.64 & 1.2 & 5.4 & 0.6 & 0.7 & -- & 1.3 \\
  H\,{\footnotesize I} & 1216 & 29.3 $\pm$ 1.9  & 37 & 28 & 18 & 41 & -- &  30\tablenotemark{g} \\
  N$_2$ LBH (4,0) & 1325 & 0.14 $\pm$ 0.10\tablenotemark{h}  & -- & -- & -- & 0.08 & 0.18 & 0.39 \\
  N$_2$ LBH (3,0) & 1354 & 0.20 $\pm$ 0.11\tablenotemark{h} & -- & -- & -- & 0.10 & 0.22 & 0.51 \\
  N$_2$ LBH (2,0) & 1383 & 0.40 $\pm$ 0.13\tablenotemark{h}  & -- & -- & -- & 0.08 & 0.18 & 0.51 \\
  N$_2$ LBH (1,1) & 1464 & 0.59 $\pm$ 0.21\tablenotemark{h}  & -- & -- & -- & -- & 0.12 &  0.29 \\
  N\,{\footnotesize I} & 1493 & 0.63 $\pm$ 0.18\tablenotemark{h}  & -- & -- & -- & 0.23 & -- & 0.68 \\
  CO 4PG (0,1) & 1597 & 1.2 $\pm$ 0.4\tablenotemark{h}  & -- & -- & -- & 0.0 & 2.0 & 1.2 \\
  CO 4PG (0,2) & 1653 & 2.9 $\pm$ 0.9\tablenotemark{h}  & -- & -- & -- & 0.0 & -- & 1.0 \\
  N$_2$ VK (7,0) & 1689 & 1.0 $\pm$ 0.7\tablenotemark{h}  & -- & -- & -- & -- & 0.21 &  0.72 \\
  \enddata
  \tablenotetext{a}{Quoted error bars are 1$\sigma$, while the upper limits are
    3$\sigma$}
  \tablenotetext{b}{\citet{Summersetal97}}
  \tablenotetext{c}{\citet{Sternetal08}}
  \tablenotetext{d}{\citet{Youngetal08}}
  \tablenotetext{e}{\citet{Stevensetal13}}
  \tablenotetext{f}{\citet{Jain:Bhardwaj15}}
  \tablenotetext{g}{In the absence of H in our AURIC model, we report
    here the value predicted by \citet{Gladstoneetal15}}
  \tablenotetext{h}{This spectral feature is a blend of multiple
    emission lines/bands. We report the total intensity of the feature,
    as determined by a Gaussian fit.}
\end{deluxetable*}

The situation is more complicated in the FUV, as Pluto's atmosphere
transitions from completely opaque for wavelengths below 1400\AA\ to $\tau<1$
for $\lambda >$1540\AA. Sunlight, reflected from the surface, overwhelms the
faint airglow emissions at wavelengths greater than
$\sim$1400\AA. Section~\ref{sec:surf} describes how we model the reflected
sunlight and subtract it from the data. However, there appears to be an
additional source of absorption in Pluto's atmosphere between
1500\AA$< \lambda < $1580\AA\ that is missing from our model. This results in
an over-subtraction of the solar spectrum, as can clearly be seen in
Figure~\ref{fig:fuv_spec}. We therefore exclude this region from all
subsequent analysis.

In addition, at the spectral resolution of Alice, emissions from the N$_2$ LBH
bands, the CO 4PG bands, N\,{\footnotesize I}, and the N$_2$ Vegard-Kaplan
(VK) bands (in blue, green, purple, and red, respectively) are significantly
blended together. Since we cannot separate these components observationally
and do not have full confidence--at the level of individual spectral
features--in the relative intensities predicted by our MLR model fit to the
data, we report only the total brightness of each spectral feature.

\subsection{Hydrogen\label{sec:H}}

The bright emission line at Lyman~$\alpha$ (1216\AA) indicates that atomic
hydrogen is present in Pluto's upper atmosphere--a result of methane
photochemistry. We find a Lyman~$\alpha$ brightness of 29.3~$\pm$~1.9~R, which
matches the pre-encounter predictions by \citet{Gladstoneetal15} that relied
on the model atmosphere of \citet{Krasnopolsky:cruikshank99}. The
H\,{\footnotesize I} Lyman~$\beta$ emission line at 1026\AA\ was also
detected, though $\sim150\times$ fainter than Lyman $\alpha$. Since the
Lyman~$\alpha$ emission line is optically thick in Pluto's atmosphere, we use
the brightness, $B$ (in units of Rayleighs), of the optically thin
Lyman~$\beta$ line to estimate the hydrogen column density above the $\tau=1$
level:

\begin{equation}\label{eq:coldens}
B =10^{-6} g_{ik} N 
\end{equation}

\noindent where the ``g-factor'', $g_{ik}$, is the number of radiative
transitions per second per particle from quantum state $k$ to state
$i$. \citet{Chamberlain:Hunten87} define the g-factor as

\begin{equation}\label{eq:gfact}
g_{ik} = \frac{\pi e^2}{m_e c^2} \frac{A_{ki}}{\sum _j A_{kj}}
  \sum _j \frac{P_j \pi F_{\sun} \lambda_{jk}^2 f_{jk}}{r^2}
\end{equation}

\noindent where the subscript $j$ in the sums on the right is necessary to
account for all possible paths to/from the upper level, $k$; $r$ is the
heliocentric distance, in AU; $F_{\sun}$ is the incident solar flux (in
photons s$^{-1}$ cm$^{-2}$ \AA-1) at 1~AU (our $F_{\sun}$ is identical to that
described in detail by \citep{Youngetal18}); $A_{kj}$ is the Einstein ``A''
coefficient for the transition from state $k$ to $j$; $f_{jk}$ is the
oscillator strength for the upward transition from level $j$ to level $k$; and
$P_j$ accounts for the portioning of levels in the ground state, given
temperature $T$:

\begin{equation}\label{eq:p}
P_j = \frac{(g_j + 1) e^{\frac{-E_j}{kT}}}{\sum _j (g_j+1) e^{\frac{-E_j}{kT}}}
\end{equation}

\noindent where $g_j$ is the statistical weight of state $j$.  We find a
g-factor for Lyman~$\beta$ of
g$_{1026 \mbox{\scriptsize{\AA}}} = 2.64\times10^{-9}$ photons s$^{-1}$. This
implies a hydrogen column density of $N_H=7.7\pm1.7\times10^{13}$~cm$^{-2}$
and a LOS mixing ratio of $8.1\pm1.8\times10^{-5}$ above the $\tau=1$ altitude
of 490~km,.

\subsection{Argon}\label{sec:Ar}

Resonant scattering of the EUV solar continuum by argon produces emission
lines at 1048\AA\ and 1067\AA, well within the bandpass of Alice. Little is
known about the relative abundance of argon in Pluto's atmosphere or other
objects in the Kuiper belt. Early observations of the atmosphere of Saturn's
moon, Titan, by Voyager IRIS placed an upper limit on the mixing ratio of
argon at 6\% \citep{Courtinetal95}. This led \citet{Summersetal97} to include
argon at a constant mixing ratio of 5\% in their Pluto atmospheric models,
resulting in a predicted brightness of the {Ar\,{\footnotesize I}}~1048\AA\
line of 0.3~R. More recently, Using the \citet{Krasnopolsky:cruikshank99}
``Model 2'' atmospheric profile and an altitude-independent argon mixing ratio
of 5\%, \citet{Sternetal08} predicted 0.45~R, and more recently,
\citet{Mousisetal13} predicted a brightness of 1.3~R--levels that should be
detectable by New Horizons' Alice. However, subsequent in situ measurements of
Titan's atmosphere by the Huygens probe gas chromatograph mass spectrometer
reduced the Voyager-era upper limit on the mixing ratio of argon by more than
three orders of magnitude to just $3.39\pm0.12\times10^{-5}$
\citep{Niemannetal10}. The ultraviolet spectrograph on Cassini, UVIS, also
failed to detect any emission from argon at Titan \citep{Stevensetal11},
calling into question whether Alice would detect argon emission at Pluto. 

Pre-flyby models of Pluto's atmosphere that predicted detectable argon
emission lines generally assumed a relatively warm, well-mixed
atmosphere. Instead, New Horizons found an atmosphere that is considerably
colder (a peak temperature of 106~K at 25~km altitude, falling to a nearly
constant temperature of 68~K in the upper atmosphere)
\citep{Gladstoneetal16}. At present, there are extreme discrepancies between
various models of Pluto's atmosphere. The model atmosphere of
\citet{Youngetal18} has a very small eddy diffusion coefficient, resulting in
an atmosphere with the well-mixed portion restricted to the planetary boundary
layer (surface to 2~km). Above that, their atmosphere is in gravitational
diffusive equilibrium. In contrast, the model atmosphere of
\citet{Luspay-Kutietal17} has a much larger eddy diffusion coefficient, such
that argon does not diffusively separate until an altitude of approximately
400~km.

We do not detect either argon emission line in the Alice data. We place a
3$\sigma$ upper limit of 0.14~R on the brightness of the {Ar\,{\footnotesize
    I}}~1048\AA\ line and 0.21~R on the {Ar\,{\footnotesize I}}~1067\AA\
line. At these wavelengths, methane is the primary source of atmospheric
opacity. For a CH$_4$ absorption cross section of $3.2\times10^{-17}$~cm$^{2}$ at
1048\AA\ \citep{Kametaetal02, Chen:Wu04}, an optical depth of $\tau=1$ is
reached at a column density of $N_{CH_4}=3.1\times10^{16}$~cm$^{-2}$. Averaged
over the field of view, this occurs at an altitude of 480~km above Pluto's
surface \citep{Youngetal18}.

The g-factor for the {Ar\,{\footnotesize I}}~1048\AA\ line at a heliocentric
distance of 32.9~AU is $7.6\times10^{-11}$ photons~s$^{-1}$. Thus, to produce
our $3\sigma$ upper limit of 0.14~R of {Ar\,{\footnotesize I}}~1048\AA\
requires an Ar column density of $1.8\times10^{15}$~cm$^{-2}$ above the
$\tau=1$ level. This is roughly 6\% of the column of density of methane. For
atmospheric models that use a small eddy diffusion coefficient
(e.g. \citet{Strobel:zhu17, Youngetal18}), the Alice detection limit isn't
significant, as even if the density of argon and (molecular) nitrogen were
equal at Pluto's surface, you would still expect brightness of the
{Ar\,{\footnotesize I}}~1048\AA\ line to be $<<0.14$~R. For atmospheric models
with high eddy diffusion coefficients, such as \citet{Luspay-Kutietal17}, the
Alice results could be more physically meaningful. 

\subsection{N\,{\footnotesize I}\label{sec:NI}} 
Several multiplets from atomic nitrogen are present in the airglow
spectrum. The brightest of these occurs in the FUV at 1493\AA, although this
feature is blended with the CO fourth positive (3,1) band and the N$_2$ LBH
(3,3) band. The EUV multiplets at 1200\AA\ and 1134\AA\ are also
significant. The observed {N\,{\footnotesize I}} 1200\AA\ multiplet is
somewhat brighter than our AURIC model predicts. However, its proximity to the
much brighter {H\,{\footnotesize I}} 1216\AA\ emission line (and resulting
scattered light results) in a low signal-to-noise ratio for this
multiplet. Although this emission feature appears to be real, it is only
present at the 1$\sigma$ level of significance.

\subsection{N\,{\footnotesize II}\label{sec:N_II}}
One of the brighter features in Pluto's EUV airglow spectrum is the
{N\,{\footnotesize II}}~1085\AA\ multiplet. This multiplet is produced
primarily by the dissociative photoionization of molecular nitrogen by solar
EUV and X-ray photons via excitation of the H band of N$^+ _2$
\citep{Samsonetal91, Bishop:Feldman03}:

\begin{equation}\label{eq:n2}
N_2 + \gamma_{\lambda < 340\mbox{\scriptsize{\AA}}} \rightarrow N(^4S) + N^+ 2s2p^3 (^3D^0 )
\end{equation}

\noindent This {N\,{\footnotesize II}}~1085\AA\ multiplet was also detected at
both Triton \citep{Broadfootetal89} and Titan \citep{Stevensetal11}. The
detection here marks the first, and thus far only, detection of ions in
Pluto's atmosphere (although the in situ instruments SWAP and PEPSSI measured
ions escaping from Pluto's atmosphere \citep{Bagenaletal16}). Although this
multiplet provides a direct detection of ion production in Pluto's upper
atmosphere, because it is a consequence of the dissociation of N$_2$ rather
than the excitation of an existing ion, it cannot be used as a diagnostic of
the ambient ion density.

\subsection{N$_2$\label{sec:N2}}
The emission feature at 980\AA\ is due to the N$_2$ Carroll-Yoshino (CY)
$c{_4} ^{\prime} \, {^1 \Sigma _u ^+} $ --- $X^1 \Sigma_g ^+ $ (0,1) band--an
electronic transition. Although the CY~(0,0) band is strongly excited by
photoelectrons and its emission was predicted by \citet{Youngetal08} and
\citet{Sternetal08}, it was not detected. This is because the CY~(0,0) band is
optically thick and strongly self-absorbed. After multiple scattering, much of
the energy is ultimately radiated away via the optically thin CY~(0,1) band
\citep{Stevensetal94, Stevens01}. Both the Voyager UVS at Triton
\citep{Broadfootetal89} and the Cassini UVIS at Titan \citep{Ajelloetal07,
  Stevensetal11, Stevensetal13} detected the CY~(0,1) band but not the
CY~(0,0) band. We adapt a multiple scattering model for the CY~(0,v'') bands
used on Earth and Titan \citep{Stevensetal94, Stevens01} to the Pluto
atmospheric profiles derived from the occultation results shown in
Figure~\ref{fig:atmprofile}. Excitation rates for $c{_4} ^{\prime}$ were
calculated from AURIC and used to initialize the model. The redistribution of
photons to more optically thin bands is calculated at milliangstrom resolution
over multiple scatterings and at all altitude layers. We find that the CY(0,0)
band is optically thick and undetectable at Pluto. In contrast, the nadir
viewing CY(0,1) emission is found to be 0.4 R, which is close to what is
observed and included in Table~\ref{tab:emissions}.

In addition to the CY~(0,1) band, emissions at several of the LBH
($a^1 \Pi_g$---$X^1 \Sigma^+_g$) bands of N$_2$ are present in the spectrum at
wavelengths greater than 1300\AA\ (see Fig~\ref{fig:fuv_spec}). Among these,
the LBH~(4,0), (3,0), and (2,0) bands at 1325\AA, 1354\AA, and 1383\AA,
respectively, are predicted to be the brightest. These features are present in
the Pluto airglow spectrum, although at fairly low signal-to-noise
levels. Many of the N$_2$ LBH bands overlap emissions from the CO fourth
positive bands.

The N$_2$ Vegard-Kaplan (VK) bands ($A^3 \Sigma^+_u$--$X^1\Sigma^+_g$) should
also be present and marginally detectable in the airglow spectrum of Pluto at
wavelengths greater than 1600\AA. In the Alice bandpass, the brightest of
these should be the (7,0) band at 1689\AA. There appears to be a weak emission
feature at this location. Other bands of the VK system are either too faint or
too blended with other emissions to be clearly detected.

\subsection{Carbon Monoxide\label{sec:CO}}

Just one month prior to the New Horizons flyby of Pluto,
\citet{Lellouchetal17} observed Pluto with the ALMA interferometer. They
report the detection of CO in Pluto's atmosphere at a mole fraction of
$515\pm40$~ppm, i.e, a surface mixing ratio of $\sim5\times10^{-4}$. At this
concentration, several of the bands of the CO fourth positive system
($A^1\Pi$---$X^1\Sigma^+$) should be detectable by Alice, although they will
be blended with the N$_2$~LBH and VK bands. Almost all of the CO fourth
positive bands are optically thick, requiring careful modeling of radiative
transfer effects to extract column density from the observed brightness. Due
to the saturation of the bands of the CO fourth positive group, Alice is not
very sensitive to changes in CO column density. For example, our modeling with
AURIC suggests that doubling the surface mixing ratio of CO leads to only a
$\sim$10\% increase in the brightest CO fourth positive bands.

Given that, our model predicts the brightest CO emission features to be the
(0,1), (0,2), (0,3), and (5,1) bands at 1597\AA, 1653\AA, 1712\AA, and
1435\AA, respectively, as shown in Fig~\ref{fig:fuv_spec}. The first three of
these bands are produced by the solar C~{\footnotesize IV}~1548\AA\ emission
line exciting the nearby CO (0,0) band at 1544\AA. Due to the large optical
depth of the (0,0) band, much of this energy is radiated away via the (0,1),
(0,2), and (0,3) bands. Similarly, the solar Si~{\footnotesize IV} emission
line at 1393.8\AA\, pumps the CO (5,0) band at 1391.1\AA and because of
optical depth effects, this energy is primarily radiated away through the
(5,1) band at 1435\AA.

\subsection{Other Features \label{sec:leftovers}}

None of the features below 920\AA\ are statistically significant. In
particular, we do not detect any emission at He\,{\footnotesize I}~584\AA\ and
place a $3\sigma$ upper limit of 0.49~R on the brightness of this emission
line. Although there appears to be a 2$\sigma$ significant emission feature at
736\AA, this is a known instrumental artifact (a Lyman-$\alpha$ ghost) and not
emission from Ne\,{\footnotesize I}.

Between 1160-1180\AA, our airglow spectrum is significantly elevated above the
background level. This feature is too wide to be due to a single emission line
or band, and none of the species that have been detected in Pluto's atmosphere
emit significantly in this bandpass. Nor are we aware of any
ions/atoms/molecules that might plausibly contribute to these putative
emissions while producing no other detectable UV signature. We therefore
believe this feature is likely an artifact of or our data processing.

Similarly, there are several features in the FUV that appear to be significant
at about the 2$\sigma$ level, yet do not correspond to the wavelength or
predicted intensity of any known emissions. Examples include the feature at
1412\AA, which is both longward of the CO Fourth Positive (6,1) band and
shortward of the N$_2$ LBH (1,0) band, and the feature at 1449\AA, which could
plausibly be the CO Fourth Positive (3,0) band, except that our AURIC model
predicts the (3,0) band should be undetectably faint.

There are several other potential features in Figure~\ref{fig:fuv_spec} at
wavelengths greater than 1600\AA. Given the 1$\sigma$ statistical error bars,
several of these appear to be real. However, systematic errors introduced by
our subtraction of the solar spectrum after modeling the atmospheric
absorption and surface reflectance are likely much greater than the
statistical uncertainty. We therefore urge caution in interpreting these
features.

\section{Spatial Distribution of Airglow Emissions \label{sec:spatial}}

From the left-hand panel of Figure~\ref{fig:geom}, the geometry of the
Airglow3 observations is such that while the field of view of detector row 16
lies almost entirely on the disk of Pluto, rows 15 and 17 (zero-indexed) lie
almost entirely off the limb, spanning a range of tangent altitudes from
0-1920 km (0--1.6 Pluto radii). No airglow emissions were detected in either
of these rows. We suggest this is a consequence of gravitational diffusion and
limb darkening. Methane (CH$_4$) is the dominant absorber below
$\sim$1450\AA. Because of its low molecular weight, it has a larger scale
height than most other atmospheric species. As a result, there is considerably
less of these heavier species between New Horizons and the $\tau=1$ level in
Pluto's atmosphere along the tangential line of sight. For example, using the
\citet{Youngetal18} atmospheric model, shown in Figure~\ref{fig:atmprofile},
the column density of $N_2$ above the $\tau=1$ level at the LBH bands is
$\sim$3.5$\times$ lower for rows 15 and 17 than it is for row 16. This
decrease in apparent column density renders the already faint emission lines
below our detection threshold.

\section{Conclusions and Future Work}\label{sec:conc}

The main conclusions of our paper are as follows:

\begin{enumerate}
\item The brightness of IPM Lyman~$\alpha$ at a heliocentric distance of
  32.5~AU in the direction of Pluto ($\alpha$=18$^h$2$^m$38.7$^s$, $\delta$ =
  -14\degr 37\arcmin37\farcs2), as seen from the New Horizons spacecraft is
  $133.4\pm0.6$~R. Lyman~$\beta$ has a brightness of $0.24\pm0.02$~R, and we
  place a $3\sigma$ upper limit of 0.10~R on the brightness of He~I~584\AA.

\item Although Pluto's atmosphere is completely opaque at Lyman~$\alpha$, it
  is optically thin ($\tau_v<1$) for photons with $\lambda>1425$~\AA. These
  FUV photons can break molecular bonds and drive photolysis on the
  surface. We suggest this has important consequences for surface weathering
  and could explain why the areas on Pluto that receive the most insolation,
  averaged over its orbit, are darker and redder than the poles. 

\item  Pluto's surface reflectance between 1400-1850\AA\ is approximately
  wavelength-independent with an I/F of 0.17. This is the first measurement of
  Pluto's reflectance in the FUV.

\item We detected a new species in Pluto's atmosphere in absorption:
  methylacetylene (C$_3$H$_4$, or propyne). In our observations,
  methylacetylene has a column density of approximately $5\times10^{15}$
  cm$^{-2}$, corresponding to a column-integrated mixing ratio of
  $1.6\times10^{-6}$. This could provide an important constraint for
  photochemical models of Pluto's atmosphere.

\item We have detected airglow emissions from N$_2$, N\,{\footnotesize I},
  N\,{\footnotesize II}, H\,{\footnotesize I}, and CO in Pluto's upper
  atmosphere. Detected emissions range in brightness from a few tenths of a
  Rayleigh to $29.3\pm1.9$~R for Lyman~$\alpha$.

\item The discovery of the N\,{\footnotesize II } multiplet at 1085\AA\ is the
  first direct detection of ions in Pluto's atmosphere. However, since this
  multiplet results from the prompt emission of N\,{\footnotesize II } after
  the dissociative photoionization of N$_2$, it is not diagnostic of
  ionospheric density.

\end{enumerate}

We suggest several areas ripe for future work. First, we examined only the
most promising subset of the Alice airglow observations, selected for their
relatively long integration time and proximity to Pluto. Second, more of the
solar spectrum is absorbed by Pluto's atmosphere/surface between
1500--1580\AA\ than we can account for in our modeling. Following our
discovery of methylacetylene, we suggest that it is likely there are one or
more additional minor species in Pluto's atmosphere that have not yet been
identified. Third, in our modeling, we have neglected the physics of Pluto's
haze particles, which are complex and likely to vary with altitude. A careful
treatment of Pluto's atmospheric haze is required to improve upon our upper
limit of Pluto's surface reflectance. Finally, although our atmospheric model
produces a reasonable match to Pluto's observed airglow spectrum, there is
significant room for improvement. A careful treatment of the radiative
transfer effects of emissions and absorption by multiple hydrocarbon species
along the line of sight could yield a significantly better match to the
observations.

\acknowledgments 

Financial support for this work was provided by NASA's New Horizons project
via contracts NASW-02008 and NAS5- 97271/TaskOrder30. We thank the New
Horizons Mission team for making these observations possible. Werner Curdt
provided the high spectral-resolution solar models. We gratefully acknowledge
the solar spectroscopic data available from the LASP Interactive Solar
Irradiance Data Center (LISIRD)
\url{http://lasp.colorado.edu/lisird/}. Atmospheric profiles of $C_3H_4$ were
kindly provided by M.~L. Wong and Y.~Yung (2017, private communication) based
on their photochemical model \citep{Wongetal17}. We thank Paul Feldman for
constructive discussion and providing model spectra of the CO Fourth Positive
bands. We are grateful to the anonymous reviewer for their helpful comments
and the editorial staff for their understanding and accommodation of the delay
between receipt of the reviewer's comments and submission of the revised
manuscript.

%

\facility{New Horizons}


\appendix
\section{Data Reduction}
In this appendix, we discuss the data reduction techniques described in
Section~\ref{sec:obs} in more detail. The first data reduction step is to
correct for the dead time of the detector, for each of the individual
exposures. This correction is necessary because the detector electronics take
a finite amount of time to process each detected count, during which the
detector is ``dead'', i.e. it is insensitive to any additional counts. Thus,
each detected count is weighted by a factor of $1 / (1 - \tau C)$, where
$\tau$=18 $\mu$s is the time constant of the electronics and $C$ is the
average count rate during the exposure \citep{Sternetal08}.

After the dead time correction we then use the ``stim pixels'' to correct the
location of the spectrum in data space \citep{Sternetal08}. Unlike many
classes of detectors such as (CCDs), the Alice detector does not have any
physical pixels. Instead, when an ultraviolet photon strikes the front surface
of the MCP, it produces a photoelectron. As the front and back surfaces of the
MCP are held at an electric potential of several thousand volts, the electron
is accelerated into the pores of the microchannel plate, where it strikes the
walls, liberating more electrons. The resulting cascade produces a cloud of
$\sim$6 million electrons ($\sim$1 pC of charge) exiting the back surface of
the MCP, which then strikes the readout anode. The detector electronics
compares the times when the charge pulse is detected on one side of the
readout anode to when the signal is detected on the other side of the anode,
after traveling through a delay circuit. The difference in timing determines
how the event is mapped from physical space on the readout anode to data
space. Changes in temperature affect the resistivity of the readout anode,
which, in turn, affects the relative timing of the charge pulses. Thus, an
event that occurs at the same physical location can be mapped into a different
location in data space, depending on temperature. To correct for this, the
electronics produces artificial charge pulses at known physical locations on
opposite sides of the detector, which allows for a linear correction to the
apparent location of detected photons.

After applying the dead time correction factor and the stim pixel correction
we co-add all the data and divide by the total exposure time of 3,900 s. We
then subtract a dark countrate image from the data. The dark countrate image
was produced by summing Alice images acquired with the airglow aperture door
closed and dividing by the total integration time. In total, 10,720 seconds of
dark integration time was obtained during Active Checkout 8 in July 2014 and
10,800 seconds of dark integration time was acquired during the post-encounter
calibration campaign of July 2016. Typical dark count rates are of the order
of 0.004 counts s$^{-1}$ px$^{-1}$ and show no evidence of temporal change.


\begin{thebibliography}{}
\expandafter\ifx\csname natexlab\endcsname\relax\def\natexlab#1{#1}\fi
\providecommand{\url}[1]{\href{#1}{#1}}
\providecommand{\dodoi}[1]{doi:~\href{http://doi.org/#1}{\nolinkurl{#1}}}
\providecommand{\doeprint}[1]{\href{http://ascl.net/#1}{\nolinkurl{http://ascl.net/#1}}}
\providecommand{\doarXiv}[1]{\href{https://arxiv.org/abs/#1}{\nolinkurl{https://arxiv.org/abs/#1}}}

\bibitem[{{Ajello} {et~al.}(2007){Ajello}, {Stevens}, {Stewart}, {Larsen},
  {Esposito}, {Colwell}, {McClintock}, {Holsclaw}, {Gustin}, \&
  {Pryor}}]{Ajelloetal07}
{Ajello}, J.~M., {Stevens}, M.~H., {Stewart}, I., {et~al.} 2007, \grl, 34,
  L24204, \dodoi{10.1029/2007GL031555}

\bibitem[{{Bagenal} {et~al.}(2016){Bagenal}, {Hor{\'a}nyi}, {McComas},
  {McNutt}, {Elliott}, {Hill}, {Brown}, {Delamere}, {Kollmann}, {Krimigis},
  {Kusterer}, {Lisse}, {Mitchell}, {Piquette}, {Poppe}, {Strobel}, {Szalay},
  {Valek}, {Vandegriff}, {Weidner}, {Zirnstein}, {Stern}, {Ennico}, {Olkin},
  {Weaver}, {Young}, {Gladstone}, {Grundy}, {McKinnon}, {Moore}, {Spencer},
  {Andert}, {Andrews}, {Banks}, {Bauer}, {Bauman}, {Barnouin}, {Bedini},
  {Beisser}, {Beyer}, {Bhaskaran}, {Binzel}, {Birath}, {Bird}, {Bogan},
  {Bowman}, {Bray}, {Brozovic}, {Bryan}, {Buckley}, {Buie}, {Buratti},
  {Bushman}, {Calloway}, {Carcich}, {Cheng}, {Conard}, {Conrad}, {Cook},
  {Cruikshank}, {Custodio}, {Dalle Ore}, {Deboy}, {Dischner}, {Dumont},
  {Earle}, {Ercol}, {Ernst}, {Finley}, {Flanigan}, {Fountain}, {Freeze},
  {Greathouse}, {Green}, {Guo}, {Hahn}, {Hamilton}, {Hamilton}, {Hanley},
  {Harch}, {Hart}, {Hersman}, {Hill}, {Hinson}, {Holdridge}, {Howard},
  {Howett}, {Jackman}, {Jacobson}, {Jennings}, {Kammer}, {Kang}, {Kaufmann},
  {Kusnierkiewicz}, {Lauer}, {Lee}, {Lindstrom}, {Linscott}, {Lunsford},
  {Mallder}, {Martin}, {Mehoke}, {Mehoke}, {Melin}, {Mutchler}, {Nelson},
  {Nimmo}, {Nunez}, {Ocampo}, {Owen}, {Paetzold}, {Page}, {Parker}, {Parker},
  {Pelletier}, {Peterson}, {Pinkine}, {Porter}, {Protopapa}, {Redfern},
  {Reitsema}, {Reuter}, {Roberts}, {Robbins}, {Rogers}, {Rose}, {Runyon},
  {Retherford}, {Ryschkewitsch}, {Schenk}, {Schindhelm}, {Sepan}, {Showalter},
  {Singer}, {Soluri}, {Stanbridge}, {Steffl}, {Stryk}, {Summers}, {Tapley},
  {Taylor}, {Taylor}, {Throop}, {Tsang}, {Tyler}, {Umurhan}, {Verbiscer},
  {Versteeg}, {Vincent}, {Webbert}, {Weigle}, {White}, {Whittenburg},
  {Williams}, {Williams}, {Williams}, {Woods}, \& {Zangari}}]{Bagenaletal16}
{Bagenal}, F., {Hor{\'a}nyi}, M., {McComas}, D.~J., {et~al.} 2016, Science,
  351, aad9045, \dodoi{10.1126/science.aad9045}

\bibitem[{{Bishop} \& {Feldman}(2003)}]{Bishop:Feldman03}
{Bishop}, J., \& {Feldman}, P.~D. 2003, Journal of Geophysical Research (Space
  Physics), 108, 1243, \dodoi{10.1029/2001JA000330}

\bibitem[{{Bockel{\'e}e-Morvan} {et~al.}(2001){Bockel{\'e}e-Morvan},
  {Lellouch}, {Biver}, {Paubert}, {Bauer}, {Colom}, \&
  {Lis}}]{Bockelee-Morvanetal01}
{Bockel{\'e}e-Morvan}, D., {Lellouch}, E., {Biver}, N., {et~al.} 2001, \aap,
  377, 343, \dodoi{10.1051/0004-6361:20011040}

\bibitem[{{Broadfoot} {et~al.}(1989){Broadfoot}, {Atreya}, {Bertaux},
  {Blamont}, {Dessler}, {Donahue}, {Forrester}, {Hall}, {Herbert}, {Holberg},
  {Hunten}, {Krasnopolsky}, {Linick}, {Lunine}, {Mcconnell}, {Moos}, {Sandel},
  {Schneider}, {Shemansky}, {Smith}, {Strobel}, \& {Yelle}}]{Broadfootetal89}
{Broadfoot}, A.~L., {Atreya}, S.~K., {Bertaux}, J.~L., {et~al.} 1989, Science,
  246, 1459, \dodoi{10.1126/science.246.4936.1459}

\bibitem[{{Chamberlain} \& {Hunten}(1987)}]{Chamberlain:Hunten87}
{Chamberlain}, J.~W., \& {Hunten}, D.~M. 1987, {Theory of planetary
  atmospheres. An introduction to their physics andchemistry.}

\bibitem[{{Chen} \& {Wu}(2004)}]{Chen:Wu04}
{Chen}, F.~Z., \& {Wu}, C.~Y.~R. 2004, \jqsrt, 85, 195,
  \dodoi{10.1016/S0022-4073(03)00225-5}

\bibitem[{{Cheng} {et~al.}(2017){Cheng}, {Summers}, {Gladstone}, {Strobel},
  {Young}, {Lavvas}, {Kammer}, {Lisse}, {Parker}, {Young}, {Stern}, {Weaver},
  {Olkin}, \& {Ennico}}]{Changetal17}
{Cheng}, A.~F., {Summers}, M.~E., {Gladstone}, G.~R., {et~al.} 2017, \icarus,
  290, 112, \dodoi{10.1016/j.icarus.2017.02.024}

\bibitem[{{Courtin} {et~al.}(1995){Courtin}, {Gautier}, \&
  {McKay}}]{Courtinetal95}
{Courtin}, R., {Gautier}, D., \& {McKay}, C.~P. 1995, \icarus, 114, 144,
  \dodoi{10.1006/icar.1995.1050}

\bibitem[{{Curdt} {et~al.}(2001){Curdt}, {Brekke}, {Feldman}, {Wilhelm},
  {Dwivedi}, {Sch{\"u}hle}, \& {Lemaire}}]{Curdtetal01}
{Curdt}, W., {Brekke}, P., {Feldman}, U., {et~al.} 2001, \aap, 375, 591,
  \dodoi{10.1051/0004-6361:20010364}

\bibitem[{{Elliot} {et~al.}(1989){Elliot}, {Dunham}, {Bosh}, {Slivan}, {Young},
  {Wasserman}, \& {Millis}}]{Elliotetal89}
{Elliot}, J.~L., {Dunham}, E.~W., {Bosh}, A.~S., {et~al.} 1989, \icarus, 77,
  148, \dodoi{10.1016/0019-1035(89)90014-6}

\bibitem[{{Evans} {et~al.}(2015){Evans}, {Stevens}, {Lumpe}, {Schneider},
  {Stewart}, {Deighan}, {Jain}, {Chaffin}, {Crismani}, {Stiepen}, {McClintock},
  {Holsclaw}, {Lef{\`e}vre}, {Lo}, {Clarke}, {Eparvier}, {Thiemann},
  {Chamberlin}, {Bougher}, {Bell}, \& {Jakosky}}]{Evansetal15}
{Evans}, J.~S., {Stevens}, M.~H., {Lumpe}, J.~D., {et~al.} 2015, \grl, 42,
  9040, \dodoi{10.1002/2015GL065489}

\bibitem[{{Gladstone} {et~al.}(1996){Gladstone}, {Allen}, \&
  {Yung}}]{Gladstoneetal96}
{Gladstone}, G.~R., {Allen}, M., \& {Yung}, Y.~L. 1996, \icarus, 119, 1,
  \dodoi{10.1006/icar.1996.0001}

\bibitem[{{Gladstone} {et~al.}(2015){Gladstone}, {Pryor}, \&
  {Stern}}]{Gladstoneetal15}
{Gladstone}, G.~R., {Pryor}, W.~R., \& {Stern}, S.~A. 2015, \icarus, 246, 279,
  \dodoi{10.1016/j.icarus.2014.04.016}

\bibitem[{{Gladstone} {et~al.}(2012){Gladstone}, {Retherford}, {Egan},
  {Kaufmann}, {Miles}, {Parker}, {Horvath}, {Rojas}, {Versteeg}, {Davis},
  {Greathouse}, {Slater}, {Mukherjee}, {Steffl}, {Feldman}, {Hurley}, {Pryor},
  {Hendrix}, {Mazarico}, \& {Stern}}]{Gladstoneetal12}
{Gladstone}, G.~R., {Retherford}, K.~D., {Egan}, A.~F., {et~al.} 2012, Journal
  of Geophysical Research (Planets), 117, E00H04, \dodoi{10.1029/2011JE003913}

\bibitem[{{Gladstone} {et~al.}(2016){Gladstone}, {Stern}, {Ennico}, {Olkin},
  {Weaver}, {Young}, {Summers}, {Strobel}, {Hinson}, {Kammer}, {Parker},
  {Steffl}, {Linscott}, {Parker}, {Cheng}, {Slater}, {Versteeg}, {Greathouse},
  {Retherford}, {Throop}, {Cunningham}, {Woods}, {Singer}, {Tsang},
  {Schindhelm}, {Lisse}, {Wong}, {Yung}, {Zhu}, {Curdt}, {Lavvas}, {Young},
  {Tyler}, {Bagenal}, {Grundy}, {McKinnon}, {Moore}, {Spencer}, {Andert},
  {Andrews}, {Banks}, {Bauer}, {Bauman}, {Barnouin}, {Bedini}, {Beisser},
  {Beyer}, {Bhaskaran}, {Binzel}, {Birath}, {Bird}, {Bogan}, {Bowman}, {Bray},
  {Brozovic}, {Bryan}, {Buckley}, {Buie}, {Buratti}, {Bushman}, {Calloway},
  {Carcich}, {Conard}, {Conrad}, {Cook}, {Cruikshank}, {Custodio}, {Ore},
  {Deboy}, {Dischner}, {Dumont}, {Earle}, {Elliott}, {Ercol}, {Ernst},
  {Finley}, {Flanigan}, {Fountain}, {Freeze}, {Green}, {Guo}, {Hahn},
  {Hamilton}, {Hamilton}, {Hanley}, {Harch}, {Hart}, {Hersman}, {Hill}, {Hill},
  {Holdridge}, {Horanyi}, {Howard}, {Howett}, {Jackman}, {Jacobson},
  {Jennings}, {Kang}, {Kaufmann}, {Kollmann}, {Krimigis}, {Kusnierkiewicz},
  {Lauer}, {Lee}, {Lindstrom}, {Lunsford}, {Mallder}, {Martin}, {McComas},
  {McNutt}, {Mehoke}, {Mehoke}, {Melin}, {Mutchler}, {Nelson}, {Nimmo},
  {Nunez}, {Ocampo}, {Owen}, {Paetzold}, {Page}, {Pelletier}, {Peterson},
  {Pinkine}, {Piquette}, {Porter}, {Protopapa}, {Redfern}, {Reitsema},
  {Reuter}, {Roberts}, {Robbins}, {Rogers}, {Rose}, {Runyon}, {Ryschkewitsch},
  {Schenk}, {Sepan}, {Showalter}, {Soluri}, {Stanbridge}, {Stryk}, {Szalay},
  {Tapley}, {Taylor}, {Taylor}, {Umurhan}, {Verbiscer}, {Versteeg}, {Vincent},
  {Webbert}, {Weidner}, {Weigle}, {White}, {Whittenburg}, {Williams},
  {Williams}, {Williams}, {Zangari}, \& {Zirnstein}}]{Gladstoneetal16}
{Gladstone}, G.~R., {Stern}, S.~A., {Ennico}, K., {et~al.} 2016, Science, 351,
  aad8866, \dodoi{10.1126/science.aad8866}

\bibitem[{{Greaves} {et~al.}(2011){Greaves}, {Helling}, \&
  {Friberg}}]{Greavesetal11}
{Greaves}, J.~S., {Helling}, C., \& {Friberg}, P. 2011, \mnras, 414, L36,
  \dodoi{10.1111/j.1745-3933.2011.01052.x}

\bibitem[{{Hendrix} \& {Hansen}(2008)}]{Hendrix:hansen08}
{Hendrix}, A.~R., \& {Hansen}, C.~J. 2008, \icarus, 193, 323,
  \dodoi{10.1016/j.icarus.2007.06.030}

\bibitem[{{Hinson} {et~al.}(2017){Hinson}, {Linscott}, {Young}, {Tyler},
  {Stern}, {Beyer}, {Bird}, {Ennico}, {Gladstone}, {Olkin}, {P{\"a}tzold},
  {Schenk}, {Strobel}, {Summers}, {Weaver}, \& {Woods}}]{Hinsonetal17}
{Hinson}, D.~P., {Linscott}, I.~R., {Young}, L.~A., {et~al.} 2017, \icarus,
  290, 96, \dodoi{10.1016/j.icarus.2017.02.031}

\bibitem[{{Hubbard} {et~al.}(1988){Hubbard}, {Hunten}, {Dieters}, {Hill}, \&
  {Watson}}]{Hubbardetal88}
{Hubbard}, W.~B., {Hunten}, D.~M., {Dieters}, S.~W., {Hill}, K.~M., \&
  {Watson}, R.~D. 1988, \nat, 336, 452, \dodoi{10.1038/336452a0}

\bibitem[{{Jain} \& {Bhardwaj}(2015)}]{Jain:Bhardwaj15}
{Jain}, S.~K., \& {Bhardwaj}, A. 2015, \icarus, 246, 285,
  \dodoi{10.1016/j.icarus.2014.08.032}

\bibitem[{{Jain} {et~al.}(2015){Jain}, {Stewart}, {Schneider}, {Deighan},
  {Stiepen}, {Evans}, {Stevens}, {Chaffin}, {Crismani}, {McClintock}, {Clarke},
  {Holsclaw}, {Lo}, {Lef{\`e}vre}, {Montmessin}, {Thiemann}, {Eparvier}, \&
  {Jakosky}}]{Jainetal15}
{Jain}, S.~K., {Stewart}, A.~I.~F., {Schneider}, N.~M., {et~al.} 2015,
  Geophysical Research Letters, 42, 9023, \dodoi{10.1002/2015GL065419}

\bibitem[{Kameta {et~al.}(2002)Kameta, Kouchi, Ukai, \& Hatano}]{Kametaetal02}
Kameta, K., Kouchi, N., Ukai, M., \& Hatano, Y. 2002, Journal of Electron
  Spectroscopy and Related Phenomena, 123, 225 ,
  \dodoi{https://doi.org/10.1016/S0368-2048(02)00022-1}

\bibitem[{{Krasnopolsky}(2020)}]{Krasnopolsky20}
{Krasnopolsky}, V.~A. 2020, \icarus, 335, 113374,
  \dodoi{10.1016/j.icarus.2019.07.008}

\bibitem[{{Krasnopolsky} \& {Cruikshank}(1999)}]{Krasnopolsky:cruikshank99}
{Krasnopolsky}, V.~A., \& {Cruikshank}, D.~P. 1999, \jgr, 104, 21979,
  \dodoi{10.1029/1999JE001038}

\bibitem[{{Lellouch} {et~al.}(2011){Lellouch}, {de Bergh}, {Sicardy},
  {K{\"a}ufl}, \& {Smette}}]{Lellouchetal11}
{Lellouch}, E., {de Bergh}, C., {Sicardy}, B., {K{\"a}ufl}, H.~U., \& {Smette},
  A. 2011, \aap, 530, L4, \dodoi{10.1051/0004-6361/201116954}

\bibitem[{{Lellouch} {et~al.}(2017){Lellouch}, {Gurwell}, {Butler}, {Fouchet},
  {Lavvas}, {Strobel}, {Sicardy}, {Moullet}, {Moreno}, {Bockel{\'e}e-Morvan},
  {Biver}, {Young}, {Lis}, {Stansberry}, {Stern}, {Weaver}, {Young}, {Zhu}, \&
  {Boissier}}]{Lellouchetal17}
{Lellouch}, E., {Gurwell}, M., {Butler}, B., {et~al.} 2017, \icarus, 286, 289,
  \dodoi{10.1016/j.icarus.2016.10.013}

\bibitem[{{Li} {et~al.}(2015){Li}, {Zhang}, {Gao}, \& {Yung}}]{Lietal15}
{Li}, C., {Zhang}, X., {Gao}, P., \& {Yung}, Y. 2015, \apjl, 803, L19,
  \dodoi{10.1088/2041-8205/803/2/L19}

\bibitem[{{Luspay-Kuti} {et~al.}(2017){Luspay-Kuti}, {Mandt}, {Jessup},
  {Kammer}, {Hue}, {Hamel}, \& {Filwett}}]{Luspay-Kutietal17}
{Luspay-Kuti}, A., {Mandt}, K., {Jessup}, K.-L., {et~al.} 2017, \mnras, 472,
  104, \dodoi{10.1093/mnras/stx1362}

\bibitem[{{Mousis} {et~al.}(2013){Mousis}, {Lunine}, {Mandt}, {Schindhelm},
  {Weaver}, {Alan Stern}, {Hunter Waite}, {Gladstone}, \&
  {Moudens}}]{Mousisetal13}
{Mousis}, O., {Lunine}, J.~I., {Mandt}, K.~E., {et~al.} 2013, \icarus, 225,
  856, \dodoi{10.1016/j.icarus.2013.03.008}

\bibitem[{{Niemann} {et~al.}(2010){Niemann}, {Atreya}, {Demick}, {Gautier},
  {Haberman}, {Harpold}, {Kasprzak}, {Lunine}, {Owen}, \&
  {Raulin}}]{Niemannetal10}
{Niemann}, H.~B., {Atreya}, S.~K., {Demick}, J.~E., {et~al.} 2010, Journal of
  Geophysical Research (Planets), 115, E12006, \dodoi{10.1029/2010JE003659}

\bibitem[{{Olkin} {et~al.}(2017){Olkin}, {Spencer}, {Grundy}, {Parker},
  {Beyer}, {Schenk}, {Howett}, {Stern}, {Reuter}, {Weaver}, {Young}, {Ennico},
  {Binzel}, {Buie}, {Cook}, {Cruikshank}, {Dalle Ore}, {Earle}, {Jennings},
  {Singer}, {Linscott}, {Lunsford}, {Protopapa}, {Schmitt}, {Weigle}, \& {the
  New Horizons Science Team}}]{Olkinetal17}
{Olkin}, C.~B., {Spencer}, J.~R., {Grundy}, W.~M., {et~al.} 2017, \aj, 154,
  258, \dodoi{10.3847/1538-3881/aa965b}

\bibitem[{{Owen} {et~al.}(1993){Owen}, {Roush}, {Cruikshank}, {Elliot},
  {Young}, {de Bergh}, {Schmitt}, {Geballe}, {Brown}, \&
  {Bartholomew}}]{Owenetal93}
{Owen}, T.~C., {Roush}, T.~L., {Cruikshank}, D.~P., {et~al.} 1993, Science,
  261, 745, \dodoi{10.1126/science.261.5122.745}

\bibitem[{{Roble} \& {Hays}(1972)}]{Roble:Hays72}
{Roble}, R.~G., \& {Hays}, P.~B. 1972, \planss, 20, 1727,
  \dodoi{10.1016/0032-0633(72)90194-8}

\bibitem[{{Samson} {et~al.}(1991){Samson}, {Chung}, \& {Lee}}]{Samsonetal91}
{Samson}, J.~A.~R., {Chung}, Y., \& {Lee}, E.-M. 1991, \jcp, 95, 717,
  \dodoi{10.1063/1.461424}

\bibitem[{{Schneider} {et~al.}(2015){Schneider}, {Deighan}, {Jain}, {Stiepen},
  {Stewart}, {Larson}, {Mitchell}, {Mazelle}, {Lee}, {Lillis}, {Evans},
  {Brain}, {Stevens}, {McClintock}, {Chaffin}, {Crismani}, {Holsclaw},
  {Lefevre}, {Lo}, {Clarke}, {Montmessin}, \& {Jakosky}}]{Schneideretal15}
{Schneider}, N.~M., {Deighan}, J.~I., {Jain}, S.~K., {et~al.} 2015, Science,
  350, 0313, \dodoi{10.1126/science.aad0313}

\bibitem[{{Siegmund} {et~al.}(2000){Siegmund}, {Tremsin}, {Vallerga}, {Beetz},
  {Boerstler}, \& {Winn}}]{Siegmundetal00}
{Siegmund}, O.~H., {Tremsin}, A.~S., {Vallerga}, J.~V., {et~al.} 2000, in
  \procspie, Vol. 4140, X-Ray and Gamma-Ray Instrumentation for Astronomy XI,
  ed. K.~A. {Flanagan} \& O.~H. {Siegmund}, 188--198

\bibitem[{{Stern}(2008)}]{Sternetal08NHoverview}
{Stern}, S.~A. 2008, \ssr, 140, 3, \dodoi{10.1007/s11214-007-9295-y}

\bibitem[{{Stern} {et~al.}(2007){Stern}, {Slater}, {Scherrer}, {Stone},
  {Versteeg}, {A'Hearn}, {Bertaux}, {Feldman}, {Festou}, {Parker}, \&
  {Siegmund}}]{Sternetal07ralice}
{Stern}, S.~A., {Slater}, D.~C., {Scherrer}, J., {et~al.} 2007, \ssr, 128, 507,
  \dodoi{10.1007/s11214-006-9035-8}

\bibitem[{{Stern} {et~al.}(2008){Stern}, {Slater}, {Scherrer}, {Stone},
  {Dirks}, {Versteeg}, {Davis}, {Gladstone}, {Parker}, {Young}, \&
  {Siegmund}}]{Sternetal08}
---. 2008, Space~Sci.~Rev., 140, 155, \dodoi{10.1007/s11214-008-9407-3}

\bibitem[{{Stern} {et~al.}(2015){Stern}, {Feaga}, {Schindhelm}, {Steffl},
  {Parker}, {Feldman}, {Weaver}, {A'Hearn}, {Cook}, \&
  {Bertaux}}]{Sternetal15ralice}
{Stern}, S.~A., {Feaga}, L.~M., {Schindhelm}, E., {et~al.} 2015, \icarus, 256,
  117, \dodoi{10.1016/j.icarus.2015.04.023}

\bibitem[{{Stevens}(2001)}]{Stevens01}
{Stevens}, M.~H. 2001, \jgr, 106, 3685, \dodoi{10.1029/1999JA000329}

\bibitem[{{Stevens} {et~al.}(2013){Stevens}, {Evans}, \&
  {Gladstone}}]{Stevensetal13}
{Stevens}, M.~H., {Evans}, J.~S., \& {Gladstone}, G.~R. 2013, in The Pluto
  System on the Eve of Exploration by New Horizons: Perspectives and
  Predictions, Laurel, MD, 165

\bibitem[{{Stevens} {et~al.}(2015){Stevens}, {Evans}, {Lumpe}, {Westlake},
  {Ajello}, {Bradley}, \& {Esposito}}]{Stevensetal15}
{Stevens}, M.~H., {Evans}, J.~S., {Lumpe}, J., {et~al.} 2015, \icarus, 247,
  301, \dodoi{10.1016/j.icarus.2014.10.008}

\bibitem[{{Stevens} {et~al.}(1994){Stevens}, {Meier}, {Conway}, \&
  {Strobel}}]{Stevensetal94}
{Stevens}, M.~H., {Meier}, R.~R., {Conway}, R.~R., \& {Strobel}, D.~F. 1994,
  \jgr, 99, 417, \dodoi{10.1029/93JA01996}

\bibitem[{{Stevens} {et~al.}(2011){Stevens}, {Gustin}, {Ajello}, {Evans},
  {Meier}, {Kochenash}, {Stephan}, {Stewart}, {Esposito}, {McClintock},
  {Holsclaw}, {Bradley}, {Lewis}, \& {Heays}}]{Stevensetal11}
{Stevens}, M.~H., {Gustin}, J., {Ajello}, J.~M., {et~al.} 2011, Journal of
  Geophysical Research (Space Physics), 116, A05304,
  \dodoi{10.1029/2010JA016284}

\bibitem[{{Stevens} {et~al.}(2017){Stevens}, {Siskind}, {Evans}, {Jain},
  {Schneider}, {Deighan}, {Stewart}, {Crismani}, {Stiepen}, {Chaffin},
  {McClintock}, {Holsclaw}, {Lef{\`e}vre}, {Lo}, {Clarke}, {Montmessin}, \&
  {Jakosky}}]{Stevensetal17}
{Stevens}, M.~H., {Siskind}, D.~E., {Evans}, J.~S., {et~al.} 2017, Geophysical
  Research Letters, 44, 4709, \dodoi{10.1002/2017GL072717}

\bibitem[{{Strickland} {et~al.}(1999){Strickland}, {Bishop}, {Evans}, {Majeed},
  {Shen}, {Cox}, {Link}, \& {Huffman}}]{Stricklandetal99}
{Strickland}, D.~J., {Bishop}, J., {Evans}, J.~S., {et~al.} 1999, \jqsrt, 62,
  689, \dodoi{10.1016/S0022-4073(98)00098-3}

\bibitem[{{Strobel} \& {Zhu}(2017)}]{Strobel:zhu17}
{Strobel}, D.~F., \& {Zhu}, X. 2017, \icarus, 291, 55,
  \dodoi{10.1016/j.icarus.2017.03.013}

\bibitem[{{Summers} {et~al.}(1997){Summers}, {Strobel}, \&
  {Gladstone}}]{Summersetal97}
{Summers}, M.~E., {Strobel}, D.~F., \& {Gladstone}, G.~R. 1997, {Chemical
  Models of Pluto's Atmosphere}, ed. S.~A. {Stern} \& D.~J. {Tholen}, 391

\bibitem[{{Throop} {et~al.}(2009){Throop}, {Stern}, {Parker}, {Gladstone}, \&
  {Weaver}}]{Throopetal09GV}
{Throop}, H.~B., {Stern}, S.~A., {Parker}, J.~W., {Gladstone}, G.~R., \&
  {Weaver}, H.~A. 2009, in AAS/Division for Planetary Sciences Meeting
  Abstracts, Vol.~41, AAS/Division for Planetary Sciences Meeting Abstracts
  \#41, 68.20

\bibitem[{{Tyler} {et~al.}(2008){Tyler}, {Linscott}, {Bird}, {Hinson},
  {Strobel}, {P{\"a}tzold}, {Summers}, \& {Sivaramakrishnan}}]{Tyleretal08}
{Tyler}, G.~L., {Linscott}, I.~R., {Bird}, M.~K., {et~al.} 2008, \ssr, 140,
  217, \dodoi{10.1007/s11214-007-9302-3}

\bibitem[{{Wagner} {et~al.}(1987){Wagner}, {Hapke}, \& {Wells}}]{Wagneretal87}
{Wagner}, J.~K., {Hapke}, B.~W., \& {Wells}, E.~N. 1987, \icarus, 69, 14,
  \dodoi{10.1016/0019-1035(87)90003-0}

\bibitem[{{Wong} {et~al.}(2017){Wong}, {Fan}, {Gao}, {Liang}, {Shia}, {Yung},
  {Kammer}, {Summers}, {Gladstone}, {Young}, {Olkin}, {Ennico}, {Weaver},
  {Stern}, \& {New Horizons Science Team}}]{Wongetal17}
{Wong}, M.~L., {Fan}, S., {Gao}, P., {et~al.} 2017, \icarus, 287, 110,
  \dodoi{10.1016/j.icarus.2016.09.028}

\bibitem[{{Woods} {et~al.}(2005){Woods}, {Eparvier}, {Bailey}, {Chamberlin},
  {Lean}, {Rottman}, {Solomon}, {Tobiska}, \& {Woodraska}}]{Woodsetal05}
{Woods}, T.~N., {Eparvier}, F.~G., {Bailey}, S.~M., {et~al.} 2005, Journal of
  Geophysical Research (Space Physics), 110, A01312,
  \dodoi{10.1029/2004JA010765}

\bibitem[{{Young} {et~al.}(2001){Young}, {Cook}, {Yelle}, \&
  {Young}}]{LYoungetal01}
{Young}, L.~A., {Cook}, J.~C., {Yelle}, R.~V., \& {Young}, E.~F. 2001, \icarus,
  153, 148, \dodoi{10.1006/icar.2001.6662}

\bibitem[{{Young} {et~al.}(1997){Young}, {Elliot}, {Tokunaga}, {de Bergh}, \&
  {Owen}}]{Youngetal97}
{Young}, L.~A., {Elliot}, J.~L., {Tokunaga}, A., {de Bergh}, C., \& {Owen}, T.
  1997, \icarus, 127, 258, \dodoi{10.1006/icar.1997.5709}

\bibitem[{{Young} {et~al.}(2008){Young}, {Stern}, {Weaver}, {Bagenal},
  {Binzel}, {Buratti}, {Cheng}, {Cruikshank}, {Gladstone}, {Grundy}, {Hinson},
  {Horanyi}, {Jennings}, {Linscott}, {McComas}, {McKinnon}, {McNutt}, {Moore},
  {Murchie}, {Olkin}, {Porco}, {Reitsema}, {Reuter}, {Spencer}, {Slater},
  {Strobel}, {Summers}, \& {Tyler}}]{Youngetal08}
{Young}, L.~A., {Stern}, S.~A., {Weaver}, H.~A., {et~al.} 2008, \ssr, 140, 93,
  \dodoi{10.1007/s11214-008-9462-9}

\bibitem[{Young {et~al.}(2018)Young, Kammer, Steffl, Gladstone, Summers,
  Strobel, Hinson, Stern, Weaver, Olkin, Ennico, McComas, Cheng, Gao, Lavvas,
  Linscott, Wong, Yung, Cunningham, Davis, Parker, Schindhelm, Siegmund, Stone,
  Retherford, \& Versteeg}]{Youngetal18}
Young, L.~A., Kammer, J.~A., Steffl, A.~J., {et~al.} 2018, Icarus, 300, 174 ,
  \dodoi{https://doi.org/10.1016/j.icarus.2017.09.006}

\bibitem[{{Zhang} {et~al.}(2017){Zhang}, {Strobel}, \& {Imanaka}}]{Zhangetal17}
{Zhang}, X., {Strobel}, D.~F., \& {Imanaka}, H. 2017, \nat, 551, 352,
  \dodoi{10.1038/nature24465}

\end{thebibliography}
\end{document}